\documentclass[superscriptaddress]{revtex4-2}
\usepackage{graphicx}
\usepackage{dcolumn}
\usepackage{bm}
\usepackage{bm}
\usepackage{subcaption} 
\usepackage{appendix}
\usepackage{amsmath}
\usepackage{amssymb}
\usepackage{amsfonts}
\usepackage[font={small}]{caption}

\usepackage{xcolor}
\begin{document}

\title{Signatures of Topological Magnon Edge States in THz Spectroscopy and Cavity Response}

\author{Ipsika Mohanty}
\affiliation{Institute for Theoretical Solid State Physics, RWTH Aachen University, Aachen, Germany}
\author{Johannes Knolle}
\affiliation{
TUM School of Natural Sciences, Physics Department, Garching, Germany
}
\affiliation{
Munich Center for Quantum Science and Technology (MCQST), Munich, Germany
}
\author{Silvia {Viola Kusminskiy}}
\affiliation{Institute for Theoretical Solid State Physics, RWTH Aachen University, 52074 Aachen, Germany}
\affiliation{Max Planck Institute for Science of Light, Staudtstraße 2, 91058 Erlangen, Germany}
\date{\today}

\begin{abstract}
Topological magnon insulators (TMIs) have emerged as promising platforms for low-energy spin-based information processing, due to their non-trivial bulk magnon topology and robust, chiral edge modes that support dissipationless transport. Although theoretical models predict these edge states, direct experimental detection remains challenging due to their limited sensitivity to conventional probes. In this work, we propose an all-optical pathway to detect topological magnon edge modes in ferromagnetic TMIs. Our approach harnesses parametric amplification of edge magnons via resonant electromagnetic driving, enabled by magnetoelectric coupling mechanisms. We concentrate on two-dimensional van der Waals ferromagnetic materials on the honeycomb lattice with magnonic band gaps in the terahertz (THz) range. We show that a spin-dependent effective electric dipole moment, arising from dynamic charge fluctuations and consistent with the lattice symmetry up to next-nearest-neighbor interactions, gives rise to one-photon-two-magnon processes leading to parametric amplification. On this basis, we propose a THz pump-probe spectroscopy protocol in which edge modes are selectively amplified and subsequently detected in absorption. Furthermore, we discuss the possibility of using THz cavities, enabling selective coupling to edge modes while filtering out bulk contributions. These findings establish a route for probing topological magnets and open new avenues for experimental exploration of exotic topological phenomena in magnetic quantum materials.
\end{abstract}

\maketitle

\begin{figure}
	\begin{center}
        \includegraphics[width = 0.95 \textwidth, trim={0 10 0 0},clip]{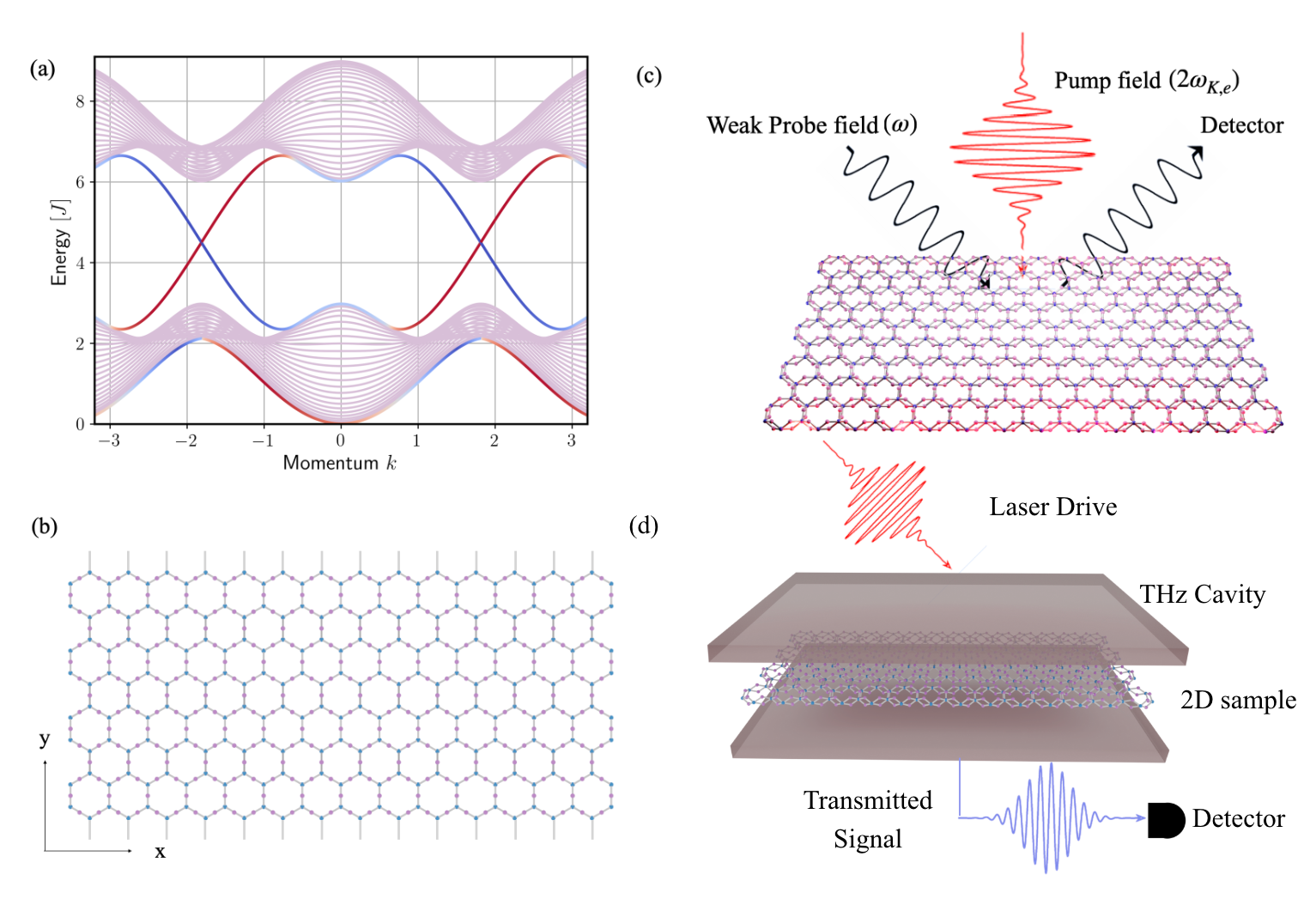}
        \captionsetup{width=0.95\linewidth}
        \captionsetup{justification=raggedright,singlelinecheck=false}
		\caption{\textbf{Magnon Band Structure and THz-Probing of a Topological Magnon Insulator} (a) Magnon band structure for a S=3/2 FM insulator on a honeycomb lattice in a ribbon geometry with a bearded boundary condition taking 24 sites in the finite direction. Pink curves indicate the bulk bands, while red and blue curves correspond to chiral edge modes. Results for n.n.n DMI $D_z=0.3J$, where $J$ is the exchange constant. (b) Honeycomb lattice extended along the $x$-direction and finite along the $y$-direction, representing a nanoribbon configuration, where the blue dots denote the transition metals (TM) and the purple dots denote halides (X) in a TM-X. The TM atoms are magnetic and form the spin lattice and the X atoms, although non-magnetic, contribute to the symmetry of the lattice. (c) Proposed pump-probe scheme for edge-mode detection, with parametric pumping at twice the edge mode frequency, $2\omega_{K,e}$ (d) Schematic of a 2D TMI coupled to a THz cavity. The cavity is driven by an external source and the transmitted signal is measured by a detector.}  
        \label{fig:Main}
	\end{center}
\end{figure}
\section{\label{Intro}Introduction}
Two-dimensional (2D) topological insulators (TI) have attracted attention due to their fundamental significance as a novel phase of matter, with prospects for applications in next-generation electronic and spintronic devices~\cite{Review_TI, Rev_TI2, Haldane, Roadmap2024, Spintronic1, Spintronic2}. Their hallmark feature, a gapped insulating bulk coexisting with robust, gapless edge states immune to back-scattering and disorder, renders them particularly attractive.  During the past few decades, the study of TIs has expanded beyond fermionic systems to encompass both interacting and non-interacting bosonic analogues~\cite{Boson_TI,Boson_TI2} including phononic~\cite{Topo_Boson_Phonon1,Topo_Boson_Phonon2,Topo_phonon}, photonic~\cite{Topo_Boson_Photon1,Topo_Boson_Photon2,Topo_Boson_light,PhysRevA.78.033834,PhysRevLett.100.013904}, mechanic~\cite{Topo_boson,Topo_mech}, and magnonic~\cite{Topo_Boson_SC,Topo_Magnon_rev} topological insulators. 

In this work, we focus on bosonic TIs based on magnons, which are quantized spin-wave excitations of magnetically ordered systems. The recent experimental realization of 2D magnets~\cite{2D_review,2D_review_2016} has motivated several theoretical proposals for realizing 2D topological magnon insulators (TMI)~\cite{honeycomb_TMI, Blatsky_DM,Theory_CrI3, honeycomb_TMI, Theory_TMI, Theory_TMI2, Mook3}, where bosonized spin-lattice models with hoppings analogous to fermionic TIs lead to a non-trivial band topology in the magnon spectrum. In ideal TMIs within the harmonic approximation, magnon edge modes can exhibit dissipationless spin transport without Joule heating, making them ideal for energy-efficient spintronic applications. TMIs also allow for amplification of edge modes~\cite{Knolle}, with possible applications as directional-amplifiers~\cite{Vittorio}. Several magnetic materials, such as $\mathrm{CrI_3}$, $\mathrm{CrXTe_3 (X=Si,Ge)}$ $\mathrm{Cu(1,3-bdc)}$, $\mathrm{Lu_2V_2O_7}$  have been identified as potential TMIs~\cite{Experiment_CrI3, Experiment_NS2, Onose_exp, zhang2024direct}.

Experimental evidence for topological magnon edge modes, however, remains challenging. Unlike electronic systems, magnons do not directly couple to charge-based probes like scanning tunneling microscopy (STM) or angle resolved photo-emission spectroscopy (ARPES), and the inherently low density of states at the edge compared to the bulk further complicates their observation via conventional scattering techniques. In addition, non-linear effects can affect the topological protection of the edge modes. Thus, evidence for non-trivial topology relies on indirect signatures such as bulk magnon bandgaps observed via inelastic neutron scattering~\cite{Experiment_CrI3, Experiment_NS2, Experiment_NS_CrI3, Crb3}, and thermal Hall responses consistent with the magnon Hall effect~\cite{Onose_exp,Mook2, Katsura_theory,Murakami,Murakami2}. Different methods have been theoretically proposed to directly probe the non-trivial topology of TMIs, e.g. by probing the bulk using optical spectroscopy~\cite{Direct_probe,shing_2026}, or by probing the edge current due to parametric amplification of magnons~\cite{Knolle}. However, the direct observation of edge modes, which would provide clear evidence for the topology, still remains a challenge. Recently, edge mode signatures have been observed using inelastic electron tunneling spectroscopy~\cite{zhang2024direct}. 

In this work, we develop a theoretical model which establishes a protocol for optically probing edge modes in ferromagnetic TMIs on a honeycomb lattice, with characteristic magnonic band gaps in the terahertz (THz) range. The protocol relies on the parametric amplification of the topological edge modes via an external electromagnetic driving field which creates magnon pairs~\cite{Knolle}. We derive the symmetry-allowed magnon-photon interactions for a ferromagnetic TMI on a honeycomb lattice~\cite{Moriya} and show that the parametric interactions necessary for amplification naturally emerge from the magnetoelectric effect, which enables electrical control of magnons in the system. We propose a pump–probe scheme where a pump beam, tuned to twice the frequency of the edge modes, selectively amplifies them while ensuring magnetic stability of the bulk. This is followed by a probe pulse that captures the spectral signatures of the amplified edge states. Alternatively, we propose to probe the edge modes by harnessing both amplification and the enhanced light-matter interaction that can be realized in a cavity. We show that a THz cavity can be used  to spectrally isolate bulk and edge contributions by selectively coupling  to the edge modes, revealing distinct edge-state signatures in the cavity response. We discuss possible realizations given the recent advances in THz cavity design \cite{Demler_Thz, Faist_2024, Daniele_Thz, THz_antennas}.

\section{\label{sec:Results} Results}
\subsection{\label{sec:Model} Model}
We consider a 2D ferromagnetic (FM) Heisenberg model on a honeycomb lattice confined to a ribbon geometry, see Fig.~\ref{fig:Main}, incorporating next-nearest-neighbor Dzyaloshinskii-Moriya interaction (DMI),
\begin{align}
    \hat{H}_{\text{FM}}=-J\sum_{\langle ij\rangle}\hat{\mathbf{S}}_{i}\cdot\hat{\mathbf{S}}_{j}+\mathbf{D}\cdot\sum_{\langle\langle ij\rangle\rangle}(\hat{\mathbf{S}}_{i}\times\hat{\mathbf{S}}_{j}) \,, \label{eq:Heis}
\end{align}
where $J>0$ denotes the strength of the nearest-neighbor Heisenberg interaction and $\mathbf{D}=D_z \hat{e}_z$ corresponds to the next-nearest-neighbor (n.n.n) DMI vector. This is a minimal model showcasing non-trivial magnon topology~\cite{honeycomb_TMI} analogous to the Haldane model~\cite{Haldane_Raghu} within the linear spin wave theory (LSWT). The system is ferromagnetically ordered out of plane when the strength of the DMI is weak compared to the Heisenberg interaction, as considered in this work. The low-energy excitations over this ordered ground state can then be obtained by implementing a Holstein-Primakoff (HP) transformation, which expresses the spins in terms of bosonic operators $\hat{a}_{k,y}$ and $\hat{b}_{k,y}$, where \textit{k} corresponds to the crystal momentum along the infinite $x$-direction and $y$ corresponds to the lattice indices along the finite $y$-direction, see Fig.~\ref{fig:Main}(b), with $a$ and $b$ denoting the two sublattices of the unit cell, see Supplementary Material (SM~\ref{AppendixA}). In the LSW approximation only quadratic terms in the bosonic Hamiltonian are retained, corresponding to a $1/S$ expansion of the spin operators. A subsequent transformation diagonalizes the finite-size system, yielding the magnon Hamiltonian \begin{equation}
\hat{H}_{FM}=\sum_{k,s}\hbar\omega_{k,s}\hat{\alpha}_{k,s}^{\dagger}\hat{\alpha}_{k,s}\,,
\end{equation}
with dispersion $\omega_{k,s}$ for the respective magnon modes $\hat{\alpha}_{k,s}$. In this system, the DMI breaks the effective time-reversal symmetry, making the system topologically non-trivial, giving rise to non-zero Chern numbers for the bulk bands. The bulk topological bands are associated with robust chiral edge modes~\cite{Mook3,BEC_Brataas}, at the harmonic level of the theory~\cite{Jonas}. The effective low-energy theory of transition metal halides (TM-X) such as $\mathrm{CrI_3}$, $\mathrm{CrB_3}$, and other complex magnetic materials such as $\mathrm{CrSiTe_3}$, $\mathrm{CrGeTe_3}$~\cite{microscopics} is known to be well captured by Eq.~\eqref{eq:Heis}, and the bulk magnon band structure predicted by the model coincides with the experimental results for $\mathrm{CrBr_3}, \mathrm{CrSiTe_3},\mathrm{CrGeTe_3}$~\cite{Crb3, Experiment_CrI3}. The characteristic frequencies of the magnons for these materials are in the range of  2-6~THz~\cite{Experiment_CrI3, Experiment_NS_CrI3}.

We consider the TMI to be coupled to an external electromagnetic (EM) field, such that the system is described by the Hamiltonian
\begin{align}\hat{H}=\hat{H}_{\text{FM}}+\hat{H}_{\text{EM}}+\hat{H}_{\text{int}}\,,
\end{align}
where $\hat{H}_{\text{FM}}$ and $\hat{H}_{\text{EM}}$ represent the free ferromagnetic and electromagnetic Hamiltonians, respectively, and $\hat{H}_{\text{int}}$ captures the interaction between the TMI and the electric field $\mathbf{E}$ and magnetic field $\mathbf{B}$. The EM field can be quantized as 
\begin{align}
     \hat{\mathbf{E}}(\textbf{r},t)=& \sum_{\textbf{p},\xi}\mathcal{E}_{\textbf{p}}\Big[\hat{c}_{\textbf{p},\xi} (t) u_{\textbf{p}}(\textbf{r}) \boldsymbol\epsilon_{\textbf{p},\xi}+\hat{c}^{\dagger}_{\textbf{p},\xi} (t) u_{\textbf{p}}^*(\textbf{r}) \boldsymbol\epsilon^*_{\textbf{p},\xi}\Big]\nonumber\\
     \hat{\mathbf{B}}(\textbf{r},t)=& i\sum_{\textbf{p},\xi}\mathcal{B}_{\textbf{p}} \Big[\hat{c}_{\textbf{p},\xi}(t) u_{\textbf{p}}(\textbf{r})(\hat{\textbf{p}} \times\boldsymbol\epsilon_{\textbf{p},\xi})-\hat{c}^{\dagger}_{\textbf{p},\xi} (t)u_{\textbf{p}}^*(\textbf{r}) (\hat{\textbf{p}} \times\boldsymbol\epsilon^*_{\textbf{p},\xi})\Big]\,,
\end{align}
where $u_{\textbf{p}}(\textbf{r})$ is the spatial mode profile normalized as $\int_V d^3ru^*_{\textbf{p}}(\textbf{r})u_{\textbf{p'}}(\textbf{r})=V\delta_{\textbf{p},\textbf{p'}}$, $\boldsymbol\epsilon_{\textbf{p},\xi}$ is the polarization vector for the polarization index $\xi$, $\hat{\textbf{p}}$ is the unit vector associated with the mode label $\textbf{p}$,  and $\hat{c}^{(\dagger)}_{\textbf{p},\xi}(t)$ correspond to the photon annihilation (creation) operators (in the Heisenberg picture) of a photon mode with frequency $\omega_\textbf{p}$. In the following, for simplicity of notation, we suppress the explicit time dependence of the photon field. The vacuum field amplitudes are $\mathcal{E}_{\textbf{p}}=\sqrt{\frac{\hbar\omega_\textbf{p}}{2 \epsilon_0 V}}$, $\mathcal{B}_\textbf{p}=\frac{1}{c}\mathcal{E}_{\textbf{p}}=\sqrt{\frac{\hbar\omega_\textbf{p} \mu_0}{2 V}}$, where $V$ is the mode volume of the EM field, and $\epsilon_0$ and $\mu_0$ are the vacuum permittivity and permeability. The free EM Hamiltonian then reduces to $H_{EM}=\sum_{\textbf{p},\xi}\hbar\omega_p\Big(\hat{c}_{\textbf{p},\xi}^\dagger \hat{c}_{\textbf{p},\xi}+1/2\Big)$. Here we work in the dipolar limit, in which the wavelength of the EM field is much larger than the magnetic sample, such that the field can be treated as spatially uniform over the sample and thus carries negligible in-plane momentum. In this regime, the coupling between EM field and TMI can be assumed to be dominated by a single EM mode close to resonance, so we restrict our formalism to a single photon field $\hat{c}$ with frequency $\omega _c$, with polarization $\boldsymbol\epsilon$. In the following, we consider results for a linearly polarized EM field. For results for a circular-polarized EM field, we refer the reader to the SM~\ref{AppendixI}.

In the dipolar limit, the magnetic field couples to the EM field resulting in an interaction Hamiltonian $\hat{H}_{\text{int}}=-\hat{\textbf{E}}\cdot\hat{\textbf{P}}-\hat{\textbf{B}}\cdot\hat{\textbf{M}}$, where $\hat{\textbf{P}}$ and $\hat{\textbf{M}}$ denote the total electric dipole and magnetic moments of the material.  The dominant interaction mechanism depends on the frequency of the driving field and the material properties. In what follows, we are interested in a regime where the EM field is tuned to match the characteristic frequencies of the magnons, which for the FM TMIs under investigation lie in the THz range. In this regime, in principle, both electric and magnetic couplings are relevant~\cite{CrI3_observation, THz_electric, THz_mag}.
We therefore analyze both the direct magnetic-dipole coupling and the electric-field-mediated interaction via a spin-dependent electric dipole moment. 

To calculate the magnon–photon coupling due to the magnetic field $\hat{H}_{\text{int}}^M=-\hat{\textbf{B}}\cdot\hat{\textbf{M}}$, we express the total magnetic moment $\mathbf{\hat{M}}=\sum_ig\mu_B\mathbf{\hat{S}}_i$ in terms of the magnon operators up to second order as 
\begin{align}
    \hat{M}^x=&\sqrt{\frac{NS}{2}}g\mu_B\sum_{s}\pi^{x}_{0,s}\hat{\alpha}_{0,s}+h.c.\,, \nonumber\\
     \hat{M}^y=&\sqrt{\frac{NS}{2}}g\mu_B\sum_{s}\pi^{y}_{0,s}\hat{\alpha}_{0,s}+h.c.\,, \nonumber\\
     \hat{M}^z=&g\mu_BNS-g\mu_B\sum_{k,s}\hat{\alpha}^\dagger_{k,s}\hat{\alpha}_{k,s}\,,
     \label{eq:magnetic moment}
\end{align} 
where $g$ is the Land\'e-factor, $\mu_B$ is the Bohr magneton, $N$ is the total number of spins, and $S=3/2$ is the spin per lattice site. The complex numerical coefficients $\boldsymbol{\pi}_{0,s}=\left(\pi^{x}_{0,s},\pi^{y}_{0,s}\right)$ are geometry-dependent and arise from the diagonalization procedure for the finite-lattice geometry (see SM~\ref{AppendixC}). Hence, to second order in the magnon operators, the coupling of the TMI to the  magnetic field is of the form
\begin{align}
    \hat{H}_{\text{int}}^M=-\hbar\sum_{s}\left(G^{M}_{\pi,0,s}\hat{\alpha}_{0,s}\hat{c}-G_{\pi,0,s}^{M}\hat{\alpha}_{0,s}\hat{c}^\dagger\right)-\hbar\sum_{k,s}G_{\pi,k,s}^M\hat{\alpha}^\dagger_{k,s}\hat{\alpha}_{k,s}\hat{c}+h.c.\,,
    \label{eq:GM}
\end{align}
where $G_{\pi,0,s}^M=i\frac{\sqrt{NS}}{2}\sqrt{\frac{\hbar\omega_c \mu_0}{2 V}}g\mu_B\boldsymbol{\pi}_{0,s}\cdot(\hat{\textbf{p}}_c\times \boldsymbol{\epsilon})$ and $G^M_{\pi,k,s}=ig\mu_B\sqrt{\frac{\hbar\omega_c \mu_0}{2 V}}$ are the coupling strengths between the TMI and the magnetic field per photon and per magnon, for in-plane and out-of-plane components of the magnetic field, respectively. 

The coupling to the electric field $\hat{H}_{\text{int}}=-\hat{\textbf{E}}\cdot\hat{\textbf{P}}$ originates from the \emph{dynamical magnetoelectric effect}. Microscopically, even when a material is non-polar, an effective electric dipole moment can arise from virtual hopping processes of electrons, either between the magnetic sites in the material or from indirect hopping via non-magnetic ligands~\cite{Adrien_d5}. This result in a spin-dependent electric dipole moment. Since we consider a centrosymmetric material, the electric dipole moment operator cannot have linear terms on the spin operators (see SM~\ref{AppendixB}). The effective electric dipole moment can be expanded in powers of the spin operators, starting from the bilinear contribution as~\cite{Moriya, Fluery_Loudon}
\begin{align}
    \hat{\textbf{P}}=\sum_{i\mu\nu}\textbf{K}_{i\mu\nu}\hat{S}_{i\mu}\hat{S}_{i\nu}+\sum_{ ij \mu\nu} \textbf{A}_{ij\mu\nu}\hat{S}_{i\mu}\hat{S}_{j\nu}+...\,,
    \label{eq:Polarization}
\end{align}
where $i,\, j$ are the site indices on the spin lattice, and $\mu,\nu = x,y,z$ are the components of the spin operators. Here, we restrict our formalism up to two-spin interactions on the lattice. Higher order spin interactions require a higher order perturbative expansion in the virtual hopping processes and therefore are expected to be smaller, and thus can be neglected. The tensors $\textbf{K}_{i \mu \nu}$ and $\textbf{A}_{ij \mu \nu}$ determine the symmetry-allowed spin-interaction terms contributing to the effective dipole moment, and can be obtained based on the symmetries of the lattice and the arrangement of the non-magnetic ligands (see SM~\ref{AppendixB}). Since the dipole moment can originate from both direct and ligand-mediated hoppings, the position of non-magnetic sites needs to be included in the symmetry analysis. 

From the expansion of the electric dipole moment in terms of the spins (Eq.~\ref{eq:Polarization}), it can be seen that the first contribution comes from quadratic spin operators on a single lattice site. Although linear-in-spin terms are forbidden, for $S\ge 1$ on-site quadratic spin interactions give rise to an effective electric dipole moment mediated by spin-orbit coupling~\cite{Symmetry_2017,Polarization_spin_2007} \footnote{On-site quadratic spin interactions originate from microscopic processes in which an electron virtually hops from the magnetic site to a neighboring non-magnetic site and back. In general due to destructive interference between the hopping processes, the net electric dipole moment is zero. However, for $S\ge1$, spin-orbit coupling can lead to a non-zero contribution. }. Based on the symmetries at the individual lattice sites, we obtain the effective spin-interaction on a single site. There are also quadratic spin interactions between nearest- and next-nearest-neighbor sites, which depend on the symmetries at the bond center. The nearest-neighbor spin interaction contributes to antisymmetric spin-induced polarization terms of the form $\hat{\textbf{S}}_i\times\hat{\textbf{S}}_j$ and $\hat{S}_{iu}^2-\hat{S}_{ju}^2$, where $i$ and $j$ denote n.n. sites and $\hat{u}$ denotes the bond-direction. The symmetry of the n.n.n. bonds allows for additional spin interaction processes including e.g. $(\hat{\textbf{S}}_i \cdot\hat{\textbf{S}}_j)\hat{\textbf{u}}$ and $(\hat{S}_{iu}^2+\hat{S}_{ju}^2)\hat{\textbf{u}}$,where $i$ and $j$ denote n.n.n sites and $\hat{u}$ denotes bond-direction (see SM~\ref{AppendixB}). 

While previous studies have explored the spin-dependent electric dipole moment for 2D materials on a ribbon geometry~\cite{Mook_2024}, the primary focus has been on polarization effects arising from nearest-neighbor spin interactions, specifically leading to the spin-current Katsura–Nagaosa–Balatsky (KNB) mechanism for the induced dipole moment $\hat{\textbf{P}}\propto(\hat{\textbf{S}}_i\times\hat{\textbf{S}}_j)\times\hat{\textbf{u}} $, where $\hat{u}$ denotes the bond direction. In terms of magnon operators, this dipole-moment term mediates scattering processes between magnons through the electric field coupling. In contrast, our work incorporates all symmetry-allowed spin-dependent dipole moment contributions up to next-nearest-neighbor interactions on the lattice, which, as we will show, give rise to new mechanisms for magnon-photon couplings. In particular, these terms are the ones leading to magnon amplification. 

Performing the HP transformation on Eq.~\eqref{eq:Polarization}, we obtain the electric dipole moment in the magnon basis up to quadratic order in magnon operators, 
\begin{align}
\hat{\mathbf{P}}_{k=0}^{(1)}&=\sum_{s}\boldsymbol{\gamma}_{0,s}\hat{\alpha}_{0,s}+h.c.\label{eq:5}\\ \hat{\mathbf{P}}_{k}^{(2)}&=\sum_{s,s'}\boldsymbol{\mu}_{k,ss'}\hat{\alpha}_{k,s}^{\dagger}\hat{\alpha}_{k,s'}+\boldsymbol{\nu}_{k,ss'}\hat{\alpha}_{k,s}\hat{\alpha}_{-k,s'}+h.c.  \,,
    \label{eq:6}
\end{align}
where $\hat{\mathbf{P}}_{k=0}^{(1)}$ and $\hat{\mathbf{P}}_{k}^{(2)}$ are the effective electric dipole moments to linear and quadratic order in the magnon operators, respectively, such that the net electric dipole moment is given as $\hat{\mathbf{P}}=\hat{\mathbf{P}}_{k=0}^{(1)}+\sum_k\hat{\mathbf{P}}_{k}^{(2)}$. The complex coefficients $\boldsymbol{\mu}_{k,ss'}$, $\boldsymbol{\nu}_{k,ss'}$, $\boldsymbol{\gamma}_{0,s}$ are material- and geometry-dependent. The interaction Hamiltonian modeling the coupling of the TMI to the electric field within the LSW approximation simplifies to
\begin{align}
\hat{H}^E_{\text{int}}&=-\hbar\sum_{s}\left(G_{\gamma,0,s}^{E}\hat{\alpha}_{0,s}\hat{c}+G_{\gamma,0,s}^{E}\hat{\alpha}_{0,s}\hat{c}^\dagger\right) -\hbar\sum_{k,s,s'} \left(\hat{c}+\hat{c}^{\dagger}\right)\left(G_{\mu,k,s,s'}^{E}\hat{\alpha}_{k,s}^{\dagger}\hat{\alpha}_{k,s'}+G_{\nu,k,s,s'}^{E}\hat{\alpha}_{k,s}\hat{\alpha}_{-k,s'}\right)+h.c.\,,
\label{eq:7}
\end{align}
where $G_{\Lambda,k,s,s'}^E=\frac{1}{\hbar}\sqrt{\frac{\hbar\omega_c}{2 \epsilon_0 V}}\boldsymbol{\Lambda}_{k,s,s'}.\boldsymbol{\epsilon}$ are the interaction strengths per photon, with $\boldsymbol{\Lambda}=\{\boldsymbol{\gamma},\boldsymbol{\mu},\boldsymbol{\nu}\}$ giving the strength of the different photon-magnon interaction processes. Eq.~\eqref{eq:7} is written in a linear-polarization basis, such that the operator $c^{(\dagger)}$ (creates) annihilates a photon with linear polarization $\boldsymbol{\epsilon}$.

\begin{figure}
    \includegraphics[width = 0.95 \linewidth]{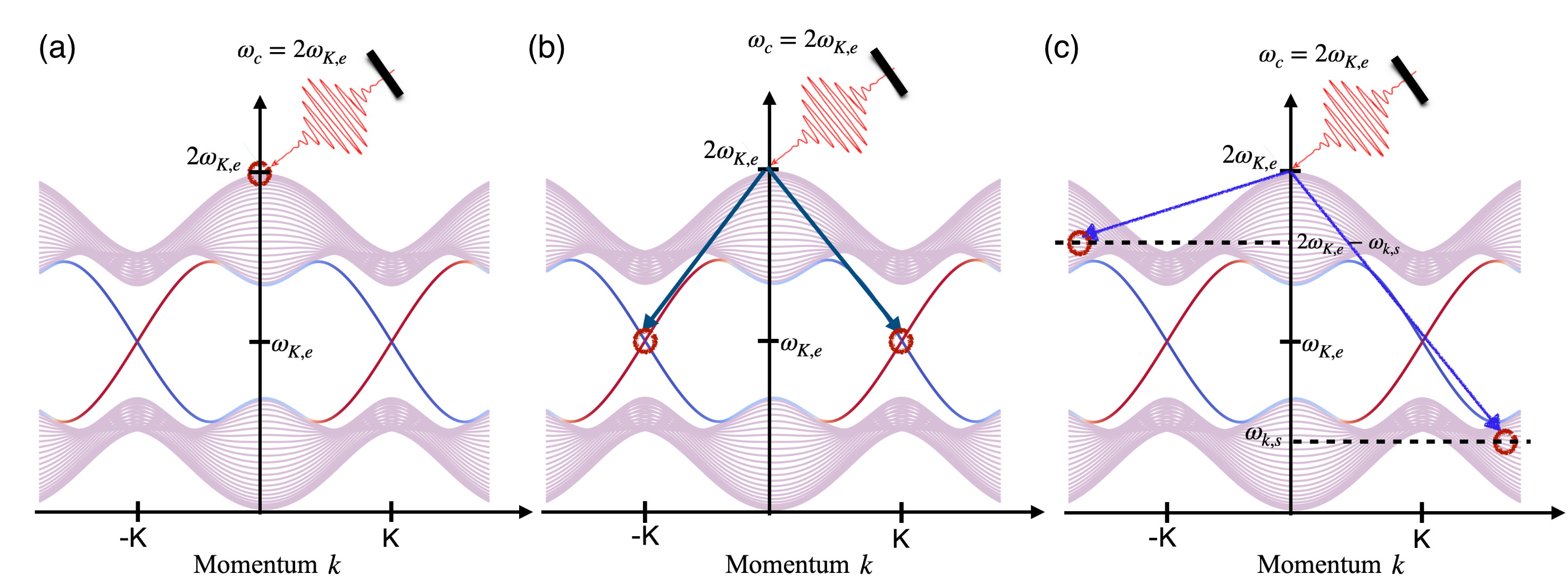}
    \captionsetup{width=0.95\linewidth}
    \captionsetup{justification=raggedright,singlelinecheck=false}
	\caption{\textbf{Symmetry-allowed photon-absorption processes } A linearly polarized EM field is incident at frequency $2\omega_{K,e}$. The FM TMI bandstructure (with frequencies shown relative to $\omega_{K,e}$) is shown together with the possible magnon-creation processes via photon absorption. The red circles denote modes in $k$-space which are resonantly excited. (a) A bulk magnon is created at frequency $2\omega_{K,e}$. (b) A pair of magnon edge modes are created at $K$ and $-K$. (c) Pair of bulk modes created at opposite momenta, satisfying the resonance condition. } 
    \label{fig:Processes}
\end{figure}

The coefficients $\boldsymbol{\mu}_{k,s,s'}$, $\boldsymbol{\nu}_{k,s,s'}$, which determine the coupling of the TMI with the electric field, depend both on the magnetic properties of the system, which control the magnitude of the electric dipole moment, as well as on the lattice geometry which dictates the $k$ and $s$ dependencies. Experimentally, extracting precise values for spin-dependent electric dipole moment contributions from one-magnon and two-magnon absorption processes is challenging. A theoretical estimate can instead be obtained using a microscopic model. For instance, for a \( \text{CrI}_3 \) monolayer, the magnitude of the two-magnon contribution to the electric dipole moment considering only n.n.n. spin interactions is estimated as (see SM~\ref{AppendixD})  \( | \boldsymbol{\nu}_{k,ss'} |\approx 50 \,\mu \text{C}/\text{m}^2 \). This value is roughly one order of magnitude smaller than the typical electric dipole moment in bulk multiferroics such as \( \text{LiCoPO}_4 \)~\cite{LiCoPo4}, but is consistent with spin-induced electric dipole moments obtained in systems such as $\mathrm{CrBr_3}, \mathrm{GaV_4S_8}$ \cite{Mook, Fumega_Lado, Nikolaev}. The relatively weak spin-dependent electric dipole moment in \( \text{CrI}_3 \) suggests a weak magnetoelectric effect compared to conventional multiferroic materials.

For the above estimates we assumed a uniform electric dipole moment in the material, however the effective electric dipole moment in the magnon basis shows a significantly larger electric dipole moment at the edges of the ribbon geometry as compared to the bulk  (see SM~\ref{AppendixC}). Additionally, one can expect that the symmetry-breaking and subsequent charge accumulation at the boundaries will lead to an enhanced electric dipole moment at the edges. Our goal is to exploit the enhanced coupling plus frequency-tuning of the drive to selectively amplify the population of edge mode magnons for optical detection. This can be achieved by the magnon-pair production arising from the coupling of the TMI to the electric field given by the couplings $G_{\nu,k,s,s'}^E$, see Eq.~\eqref{eq:7}. To activate these processes, we utilize a driving scheme resonant with twice the magnon frequency, which can lead to parametric amplification of magnon pairs. In the next section, we explore the dynamics of the TMI under such a selective driving protocol.

\begin{figure}
    \includegraphics[width = 0.95 \linewidth]{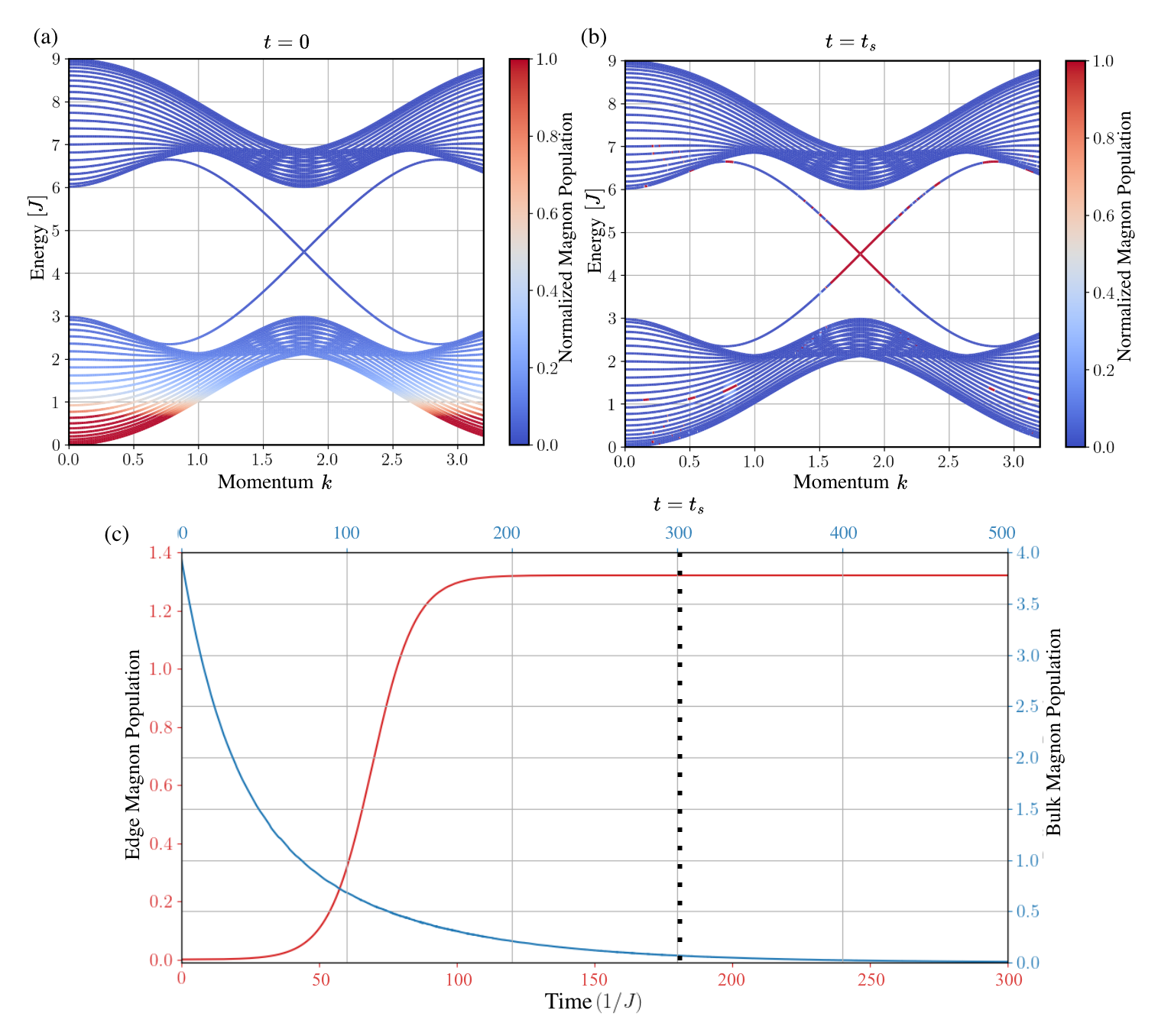}
    \captionsetup{width=0.95\linewidth}
    \captionsetup{justification=raggedright,singlelinecheck=false}
	\caption{\textbf{Edge mode amplification under driving at $2\omega_{K,e}$ } (a) Magnon band structure with normalized magnon population encoded in the colors. The first panel shows the initial state, with a magnon population in the low energy states. The second panel shows the magnon population after amplification at $t=t_s$ when the system has reached the steady state. A maximum population in the edge modes can be observed. (b) Temporal dynamics of the edge mode (red) and a bulk mode (blue) under driving at frequency $2 \omega_{K,e}$. We plot the magnon population at a $K$-point, averaged over the finite sites along the $y$-direction. The edge mode shows amplification and subsequent stabilization, while bulk modes relax. The linear and non-linear dissipation are set to $0.04\,J$ and $0.004\,J$~\cite{Jonas}, respectively. The magnitude of coupling between the drive and the edge magnons $\tilde{G}^E_{\nu.K,e,e}$ is taken as $0.15\,J$. For $\mathrm{CrI_3}$ the exchange coupling can be estimated to be  $J\approx0.5\,\mathrm{THz}$. Results are shown for a drive intensity $I=4\times 10^{13}\,\text{W}/\text{m}^2$. } 
    \label{fig:Amplification}
\end{figure}

\subsection{\label{sec:Dynamics}Dynamics}
We study the magnon dynamics of the TMI under a coherent EM drive. For a strong coherent drive, the photon operators associated with the EM mode of the incident field,  $\hat{c}$ and $\hat{c}^\dagger$ can be replaced by their expectation values, $ce^{-i\omega_ct}$ and $c^*e^{i\omega_ct}$, respectively, where $\omega_c$ is the frequency of the drive and c denotes the complex amplitude of the coherent field. In this limit, the EM field acts as a classical drive of the TMI, with field expectation value $\langle E_c\rangle=\mathcal{E}_c(ce^{-i\omega_ct}+c^*e^{i\omega_ct})$, which scales with mean photon number $n_c=|c|^2$ as $E_c=\sqrt{n_c}\mathcal{E}_c$. The strength of the drive is controlled experimentally by the intensity $I$ of the incident field, which is related to the field amplitude as $I=\frac{1}{2}c\epsilon_0E_c^2$. In presence of the classical drive, the light-matter coupling $G$ is modified as $\tilde{G}=G|c|=G\sqrt{n_c}=G\sqrt{2I/(c\epsilon_0\mathcal{E}_c^2)}$, i.e., $\tilde{G}\propto\sqrt{I}$. This allows the coupling to be directly tuned by the intensity of the drive.

To selectively probe the edge modes, we consider an in-plane polarized drive with an electric field linearly polarized along the $y$-direction, as this configuration maximizes the coupling to pairs of magnon edge modes for the bearded configuration (see SM \ref{AppendixC}). We further consider the frequency of the drive to be resonant with the creation of pairs of magnon edge modes, i.e. $\omega_c=\omega_{K,e}+\omega_{-K,e}=2\omega_{K,e}$, where $\omega_{\pm K,e}$ is the frequency of the edge mode at the degeneracy points, $k=K$. Since the magnon-drive coupling strength is small compared to the frequency of the magnons and the drive, we can use the rotating wave approximation (RWA), neglecting the fast rotating terms. So, only processes satisfying the frequency-matching conditions have a significant contribution to the dynamics. In particular, one-magnon excitations and magnon-pair creation at this frequency have significant contributions, while magnon-scattering processes induced by the drive are far detuned and can be neglected (see Fig.~\ref{fig:Processes}). The Hamiltonian for the TMI subjected to a coherent drive at frequency $\omega_c$, under the RWA in the interaction picture, reduces to (see SM~\ref{AppendixC})
\begin{align}
    \hat{H}_{int}&=-\hbar\sum_{s}(\tilde{G}^M_{\pi,0,s}+\tilde{G}_{\gamma,0,s}^{E})\hat{\alpha}_{o,s}^\dagger c e^{i(\omega_{0,s}-\omega_c)t}-\hbar\sum_{k,s,s'} \tilde{G}_{\nu,k,s,s'}^{E}\hat{\alpha}_{k,s}^\dagger\hat{\alpha}_{-k,s'}^\dagger c e^{i(\omega_{k,s}+\omega_{k,s'}-\omega_c)t}+h.c.\,,
    \label{Eq:two_magnon}
\end{align}
where $\tilde{G}_{\Lambda,k,s,s'}^E=\frac{1}{\hbar}\sqrt{\frac{2I}{c\epsilon_0}}\boldsymbol{\Lambda}_{k,s,s'}.\boldsymbol{\epsilon}$ and  $\tilde{G}_{\pi,0,s}^M=i\frac{\sqrt{NS}}{2}\sqrt{\frac{2I \mu_0}{c}}g\mu_B\boldsymbol{\pi}_{0,s}\cdot(\hat{\textbf{p}}_c\times \boldsymbol{\epsilon})$ are the magnon couplings to the electric and magnetic field of the drive, respectively. One-magnon processes contribute significantly when the frequency of a single magnon mode $\hat{\alpha}_{0,s}$, is resonant with the drive frequency, i.e., $\omega_{0,s} \approx \omega_c$. Similarly, the magnon-pair creation processes contribute to the dynamics when $\omega_{k,s}+\omega_{-k,s} \approx\omega_c$. It is the latter process that we want to select through the drive.

In order to obtain the dynamics of the coupled system, it is convenient to move to a rotating frame defined by $\hat{\alpha}_{k,s}(t)=\hat{\alpha}_{k,s}(t)e^{-i\omega_ct/2}$, which removes the explicit time-dependence of the parametric terms near resonance. In this frame, the Heisenberg equations of motion in terms of the classical amplitudes of the magnon and photon operators reduce to
\begin{align}
    \frac{\partial \langle \hat{\alpha}_{k,s}\rangle}{\partial t}=&\left(-i\Delta_{k,s}-\frac{\zeta}{2}-\frac{\eta}{2}|\langle\hat{\alpha}_{-k,s}\rangle|^2\right)\langle\hat{\alpha}_{k,s}\rangle+i\sum_{s'} \tilde{G}_{\nu,k,s,s'}^E  \langle\hat{\alpha}_{-k,s'}^{\dagger}\rangle  c +i(\tilde{G}^E_{\gamma,0,s}+\tilde{G}^M_{\pi,0,s})  ce^{-i\omega_ct/2}
\delta_{k,0}\,,\\ \nonumber
 \frac{\partial \langle \hat{\alpha}_{-k,s}^{\dagger}\rangle}{\partial t}=&\left(i\Delta_{-k,s}-\frac{\zeta}{2}-\frac{\eta}{2}\eta|\langle\hat{\alpha}_{k,s}\rangle|^2\right)\langle\hat{\alpha}_{-k,s}^{\dagger}\rangle-i\sum_{s'} \tilde{G}_{\nu,k,s,s'}^E   c^* \langle\hat{\alpha}_{k,s'}\rangle -i(\tilde{G}^E_{\gamma,0,s}-\tilde{G}^M_{\pi,0,s})c^* e^{i\omega_ct/2}\delta_{k,0}\,,
\end{align}
where $\Delta_{k,s}=\omega_{k,s}-\omega_c/2$ is the detuning and we have neglected correlations, i.e. $\langle\alpha_{k,s}\alpha_{-k,s'}\rangle \approx \langle\alpha_{k,s}\rangle \langle\alpha_{-k,s'}\rangle$. The equations of motion include linear ($\zeta$) and nonlinear ($\eta$) magnon damping terms, originating from higher order magnon-scattering processes such as three- and four-particle processes
~\cite{Nonlinear_damp,Gardiner,Non_linear_FM}. The processes responsible for linear damping lead to both broadening of magnon lifetimes and renormalization of the band structure~\cite{Jonas}. However, since the frequency renormalization is significantly smaller than the magnon frequencies, we neglect these contributions and only include the dissipative term. The non-linear processes, which  can arise either due to magnon-magnon interactions within the model or due to two-magnon losses to a bath~\cite{Gardiner}, lead to an amplitude dependent damping, which is essential to saturate the amplified modes such that the system reaches a steady state~\cite{Knolle}. 

Solving the above EOMs, we obtain the steady state dynamics of the system, neglecting the effect of fluctuations and noise in the dynamics. Under strong coherent driving, it is expected that when the incident power of the external driving field crosses a threshold, the magnon population $|\langle\hat{\alpha}_{k,s}\rangle|^2$ increases exponentially for modes which satisfy $\omega_c\approx\omega_{k,s}+\omega_{-k,s}$. This exponential increase is stabilized by the non-linear damping to yield an amplified steady-state occupation, where the time required for the system to stabilize depends on the interplay between dissipation and driving~\cite{Knolle, Vittorio}. We verify the amplification and subsequent stabilization of magnon modes by numerically simulating the driven system on a ribbon geometry with 24 finite lattice sites in the $y$-direction, which is sufficient to obtain well localized edge modes. We initialize the system close to its ground state, with a small magnon population equivalent to a thermal occupation at temperature $T=10^{-3}\,J/k_B$. Initializing the system at large occupations is avoided since the linear spin wave theory, which is essential to ensure topological protection of edge modes~\cite{Jonas}, breaks down. Upon application of an external electric field polarized in the $y$-direction, at frequency $\omega_c=2\omega_{K,e}$ i.e., resonant with twice the edge mode frequency, we observe an amplification of the magnon occupation in the edge modes. As expected, the nonlinear damping stabilizes this amplification, yielding a steady-state response as can be seen in Fig.~\ref{fig:Amplification} (a,b). We also see that the bulk modes are not amplified under the driving at $\omega_c=2\omega_{K,e}$, which is ensured due to the detuning of the bulk modes from the external drive, and also partly maintained because of the weak interaction of these modes with external fields due to the low bulk electric dipole moment as compared to the edge electric dipole moment (See SM~\ref{AppendixC}).  In addition, we see from the magnon dispersion in Fig.~\ref{fig:Processes}, that a few bulk modes are also resonant with the drive, $\omega_c=\omega_{0,s}$ leading to single-magnon bulk processes, which can result in unwanted spectral signatures. 

\begin{figure*}
	\begin{center}
        \includegraphics[width = 0.95 \textwidth]{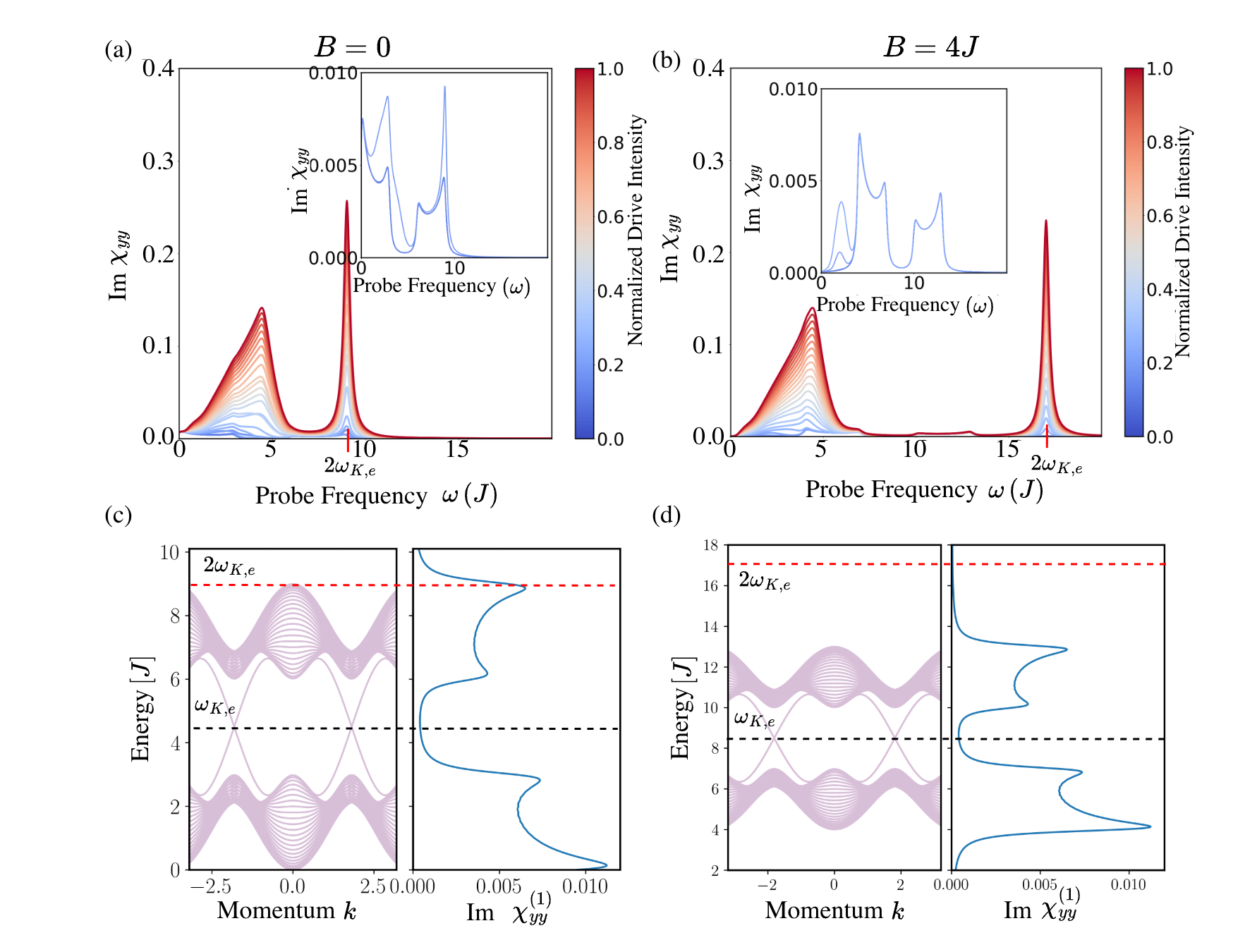}
        \captionsetup{width=0.95\linewidth}
        \captionsetup{justification=raggedright,singlelinecheck=false}
		\caption{\textbf{Response of the TMI under a resonant pump drive at $2\omega_{K,e}$ for increasing pump-intensity} A pump laser is incident at $2\omega_{K,e}$ and a probe laser after time $t_s$. The susceptibility is obtained as a function of the probe laser frequency for varying intensity of the pump-laser. The color gradient, blue (low intensity) to red (high intensity), shows the normalized pump intensity corresponding to drive-magnon edge mode coupling strengths $\tilde{G}_{\nu,K,e,e}$ in the range $0.01-0.25J$. This is equivalent to varying the pump-intensity in the range $10^{12}-10^{14}\,\text{W/m}^2$. (a) Without a bias magnetic field, the resonance frequency is $2\omega_{K,e}=9J$. The peak at the lower frequency range is due to scattering between two magnon modes and the sharp peak at $9J$ is due to two-magnon creation with energy $4.5J$ each. Comparing with the band structure, we see that this is a signature of the edge mode. The inset shows the response function for lower pump intensities, corresponding to coupling strengths up to $0.08J$, which shows spectral signatures between the two peaks, corresponding to the one-magnon response. (b) With a bias magnetic field, $B_0= 4J$. The pump frequency is $2\omega_{K,e}=9J+2B_0=17J$. The sharp peak at higher frequency is shifted by twice the magnetic field and is the response from the amplified edge mode. The lower frequency response is due to drive induced magnon scattering. The response in between the peaks comes from the one-magnon processes ( also shown in the inset as dominating response for lower driving intensities). The band structure and the corresponding susceptibility for one magnon processes (c) without and (d) with the magnetic field show a shift in energy. The estimated value for $J$ is $0.5\,\mathrm{THz}$, which gives a maximum intensity of the order $\approx10^{14}\,\text{W}/\text{m}^2$ (see SM~\ref{AppendixD}).}
        \label{fig:Susceptibility}
	\end{center}
\end{figure*}

In order to avoid single-magnon contributions from the bulk, a static magnetic field can be applied to the TMI along the direction of magnetization i.e., along the $z$-direction. This leads to a shift in the magnon modes by a constant proportional to the magnetic field strength $B_0$ (Fig.~\ref{fig:Susceptibility} d), whereas the two-magnon frequency is shifted by $2B_0$, ensuring that there are no bulk modes at $k=0$ which would satisfy the frequency matching condition with the drive. The shift in the spectrum also shifts the three magnon processes (e.g. $\alpha_{0,s}\alpha^\dagger_{k,s}\alpha^\dagger_{
-k,s'}$)  away from resonance. These terms could potentially lead to coherent amplification of edge modes or even bulk modes, which has been neglected in our protocol by treating three magnon processes as dissipative channels. When the magnon-drive coupling is comparable to the magnon frequencies, the RWA breaks down and other bulk modes can also be excited by pair-creation, which can lead to an increased effective temperature of the material, eventually destabilizing the magnetic order.

Finally, we note that the coupling to an external field leads to a splitting of the edge modes at the degeneracy point $K$ (SM~\ref{AppendixE}). This requires the drive linewidth to be larger than the splitting of the edge modes, in order to simultaneously address the two edge modes participating in the parametric pair-creation process. At the same time, the energy separation between edge and bulk modes should be discernible, imposing an upper boundary on the drive linewidth. If the drive linewidth is broader than this gap, the drive can induce significant pair-creation of bulk magnons, leading to demagnetizing effects. Hence the intensity, frequency, and linewidth of the external drive have an optimal window for edge-selective amplification.

\subsection{\label{sec:Probe}Probing the edge modes}
In the previous section, we studied the dynamics of the TMI under an external electromagnetic driving. In this section, we discuss two means for direct detection of edge modes: a free-space pump-probe scheme, and spectral signatures in the response of a driven THz cavity.

\subsubsection{\label{sec:Pump Probe}Pump-probe Spectroscopy}
To extract observable signatures of the edge modes, we propose a pump–probe spectroscopy protocol that exploits parametric amplification under resonant driving. A linearly polarized coherent THz laser field, with electric field along the $y$-direction pumps the system at $\omega_c = 2\omega_{K,e}$. A weak probe pulse is then incident on the material after the system has reached its non-equilibrium steady state under pumping. The response of the material can be measured by the absorptive part of the susceptibility given by~\cite{Mook_2024} 
\begin{align}
    \text{Im}\chi_{yy}(\omega) =\text{Im} \chi_{yy}^{(1)}(\omega)+\text{Im} \chi_{yy}^{(2)}(\omega)\,,
\end{align}
where $\text{Im}\chi_{yy}(\omega)$ measures the response of the system in the $y$-direction for incident light polarized along the $y$-direction. $\text{Im} \chi_{yy}^{(1)}(\omega)$ and $\text{Im} \chi_{yy}^{(2)}(\omega)$ are the susceptibilities due to one-magnon and two-magnon processes, respectively obtained in the non-equilibrium steady state (see Methods~\ref{Methods_2}).

At low pump intensities, the response is dominated by one-magnon processes (Fig.~\ref{fig:Susceptibility} (a,b)), $\text{Im} \chi_{yy}^{(1)}(\omega)$. These depend on the magnon density of states and not on the magnon population, and therefore their contribution remains constant independently of the pump intensity (Methods~\ref{Methods_2}). $\text{Im} \chi_{yy}^{(1)}(\omega)$ shifts uniformly when a magnetic field is applied, reflecting the corresponding shift of the magnon spectrum in the presence of a static magnetic field (see, Fig.~\ref{fig:Susceptibility}(c,d)).

Above a critical pump-power threshold, the system undergoes exponential amplification of the edge-mode population, as described in the previous section. At these pump powers, two-magnon processes dominate the response (see Fig.~\ref{fig:Susceptibility}). At low frequencies, one can observe signatures of scattering between bulk magnons. At a frequency resonant with $2\omega_{K,e}$, we see a sharp peak, which is proportional to the amplified edge mode population and hence serves as a direct signature for the edge modes. The intensity of the response increases with increasing pump intensity. However, the laser intensity is limited, as excessively high intensities may lead to demagnetization of the material due to bulk excitation. Therefore, the pump laser has to be operated in an optimal window to obtain a signature of the edge modes. In the SM~\ref{Appendix_Demag} we show that for the intensities shown in Fig.~\ref{fig:Susceptibility} demagnetization effects can be neglected. 

At intermediate pump intensities, just above the threshold, at $2\omega_{K,e}$ one can expect signatures of both bulk magnons from one-magnon processes, and from the edge mode arising from magnon pairs, see inset in Fig.~\ref{fig:Susceptibility}(a). To distinguish these signatures, we study the susceptibility under a magnetic field, applied along $z$. This shifts the whole spectrum by $B_0$ and hence the pump-frequency needs to be set at $2(\omega_{K,e}+B_0)$ to resonantly excite edge mode pairs.In absence of the external magnetic field, the resonant frequency $2\omega_{K,e}$ overlaps with the single magnon bulk band (see, Fig.~\ref{fig:Susceptibility}(d)). In presence of the magnetic field, the single magnon bulk band shifts by $B_0$, while the resonant frequency is shifted by $2B_0$, leading to the resonant frequency to be separated from the bulk band (see Fig.~\ref{fig:Susceptibility}(d)). Fig.~\ref{fig:Susceptibility}(b) shows that the one-magnon response is shifted by $B_0$, while two-magnon response is shifted by $2B_0$ and thus can be easily distinguished. 

In Fig.~\ref{fig:Susceptibility} we show results for parameters corresponding to $\mathrm{CrI_3}$ for pump intensities in the range $10^{12}\,\text{W}/\text{m}^2$ to $10^{14}\,\text{W}/\text{m}^2$. The threshold pump intensity above which the two magnon response dominates depends on the magnon dissipation rates. We set a linear magnon dissipation rate of $\zeta=0.04J\sim0.02\text{THz}$~\cite{Jonas}, consistent with estimated values for Gilbert damping in the material~\cite{Gilbert_const}. Since the non-linear dissipation originates from a higher order processes~\cite{Gardiner}, we assume it to be suppressed relative to linear dissipation, $\eta=0.004J\sim0.002\,\text{THz}$ (see also~\cite{non_gilbert_damp_2026}). For these dissipation rates, the threshold drive-magnon coupling required to observe the pair-creation of magnons is $\tilde{G}^E_{\nu,k,s,s'}>0.04J\sim0.02\,\text{THz}$, corresponding to an intensity of order $10^{13}\,\text{W}/\text{m}^2$. Currently, such high peak intensities can be achieved experimentally using ultrashort laser pulses with fluence of the order of $0.1\,\mathrm{mJ/cm}^2$, with a pulse duration of roughly $100\,\mathrm{fs}$~\cite{Prineha_Magnetic_phase}. The onset of magnon amplification is determined by the ratio of the exchange interaction $J$ and the magnon dissipation rate $\zeta$. For the parameters considered above, a pulse duration of $10\,\mathrm{ps}$ is required for the onset of amplification, and reaching the steady state requires a time longer than $0.1\,\mathrm{ns}$. Generating such a pulse lasting for $0.1\,\mathrm{ns}$ is technologically challenging. Alternatively, an average intensity of the order $10^{13}\,\text{W}/\text{m}^2$ needs to be maintained, which at a repetition rate of $20\,\text{MHz}$ requires a fluence $\sim10^4\,\text{mJ/cm}^2$. At such high average intensities the stability of the material may become a concern, requiring an efficient heat dissipation mechanism. Thus, the maximum intensity is limited and ultimately depends on the material properties and the possibility of devising heat sinks. Whereas challenging, it has been demonstrated that efficient power dissipation can be engineered by an appropriate choice of a substrate ~\cite{Huebl_spin_transport}. Moreover, it is essential to ensure that the linewidth of the pump-laser is broader than the edge-mode splitting due to the pump-laser (see SM~\ref{AppendixE}), here $0.025\text{THz}$. The value of the magnetic field assumed in Fig.~\ref{fig:Susceptibility} is only for depiction purposes. A magnetic field of strength $B_0=10\,\text{T}$, which would lead to a Zeeman shift equivalent to $0.28\text{THz}$ is sufficient to resolve the one-magnon bulk response and the response from the edge modes. 

The threshold pump intensities required for magnon edge amplification depends on the polarization  of the incident field, as we show in SM~\ref{AppendixI}. For the considered ribbon geometry, the coupling of the TMI to left circularly polarized (LCP) light is larger than the coupling to linearly or right circularly polarized (RCP) light. We estimated the threshold driving intensity for LCP light to be $\sim10^{11}\text{W/m}^2$, which is one order of magnitude lower as compared to the threshold intensity required for linearly polarized field. In addition, the requirement on the threshold pump-driving intensity and linewidth is also sensitive to the boundary conditions. In SM~\ref{AppendixG} we show  that for zigzag boundary conditions strong demagnetization effects occur under same driving conditions as used for bearded boundaries. For zigzag boundary conditions, the edge modes are close in frequency to the bulk bands, making the edge mode resolution difficult. This also leads to unwanted bulk excitations under parametric driving of edge modes. The excitation of bulk modes can be avoided by applying a strong magnetic field along the $z$-direction, which leads to spectral separation of the magnon edge-pair from the single magnon bulk band as discussed above.

\begin{figure*}
	\begin{center}
        \includegraphics[width = 0.95 \textwidth, trim={0 0 0 0},clip]{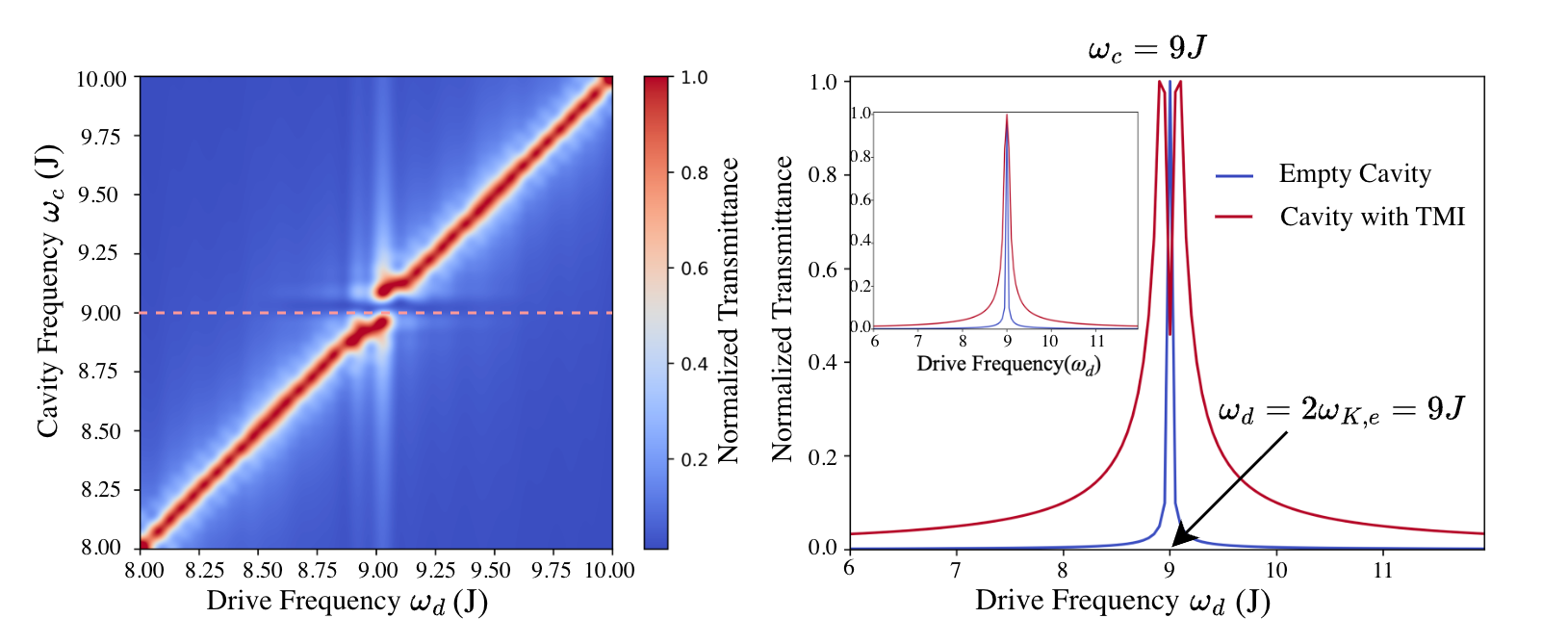}
\captionsetup{width=0.95\linewidth}
        \captionsetup{justification=raggedright,singlelinecheck=false}
		\caption{\textbf{Cavity spectrum}  (a) Normalized transmittance for a cavity coupled to the TMI, with cavity dissipation rate $\kappa=0.01J$, dissipation of the cavity due to coupling to external ports $\kappa_{ex}=0.005J$, and input field amplitude $\langle \hat{c}_{in} \rangle=10$. The linear and non-linear magnon dissipation parameters considered are $\zeta=0.04J$ and  $\eta=0,004J$, respectively and the coupling to the edge mode at the $K$-point is taken to be $\tilde{\mathcal{G}}^{E}_{\nu,K,e,e}=0.05J$. A hybridization between cavity and a discrete edge pair can be observed at the two-magnon resonant frequency $9J$. The 2D plot is smoothened using Lanczos interpolation.(b) A cross-section of the transmittance plot, with $\omega_{c}=9J$, shows broadening of the linewidth and a dip at the resonant frequency, which is a signature of the edge-mode resonance. In the inset, the cavity transmittance is plotted without including the coupling to the edge modes. In this case we see only the broadening of the linewidth.}
        \label{fig:Cavity}
	\end{center}
\end{figure*}

\subsubsection{\label{sec:Cavity Probe}Cavity as a probe}
While parametric amplification under direct driving provides a viable route to accessing edge-mode signatures, it typically requires high input powers due to the weak effective electric dipole moment of the material. To overcome this limitation, we propose an alternative approach using a THz cavity as a sensitive probe of edge magnons, see Fig.~\ref{fig:Main}. A cavity enhances the coupling strength due to mode confinement (the vacuum fluctuations of the EM field scale as $1/\sqrt{V_{cav}}$, where $V_{cav}$ is the cavity volume). Further, it acts as a frequency filter with a bandwidth determined by its quality factor, enabling selective probing of the edge-magnon pairs. 

We consider the topological magnon insulator placed within a THz cavity, positioned such that the electric field antinode of the cavity mode overlaps with the magnetic sample \cite{Thz,Demler_Thz}, in order to maximize the electric dipole coupling to the magnons. The cavity is driven by a coherent external laser field. In contrast to the free space pump-probe scheme discussed above, the photon field dynamics now plays a significant role. The dynamics of the full system is described by equations of motion that incorporate the cavity field and the magnon amplitudes (see, Eq.~\ref{eq:cav}). We numerically solve the full coupled classical dynamics of the coherent photon and magnon amplitudes, neglecting fluctuations and correlations. The cavity transmission spectrum is then obtained using input–output theory (see Methods~\ref{Methods_3}).

Fig.~\ref{fig:Cavity}(a) shows the cavity transmission spectrum when the cavity is tuned close to the parametric resonance condition $\omega_c\approx2\omega_{K,e}$, corresponding to the frequency of the edge-magnon pair. The laser driving the cavity is also tuned close to this resonance frequency, $\omega_d\approx2\omega_{K,e}$. The coupling between the cavity and the magnon pair leads to a splitting in the transmission spectrum. Fig.~\ref{fig:Cavity}(b) shows a cross-section at fixed cavity frequency $\omega_c=2\omega_{K,e}$ as a function of the driving frequency $\omega_d$.  A broadening of the linewidth is observed, due to the coupling of the cavity to the bulk-magnon-pairs which are also resonant at $2\omega_{K,e}$, which can be seen from the two-magnon density of states, see Fig.~\ref{fig:demag}. The bulk magnon-pairs effectively act as dissipative channels for the cavity, leading to broadening of the cavity linewidth. In addition to the broadening, a distinct dip can be observed in the transmission spectrum, which can be attributed to the enhanced coupling between the cavity and the edge magnon-pairs. As discussed before, the electric dipole coupling to the edge modes is significantly stronger than the bulk. In absence of this coupling, we obtain only broadening of the linewidth of the cavity spectrum as shown in the inset of Fig.~\ref{fig:Cavity}.  Thus the additional dip in the transmittance is the signature of the edge modes in the TMI. Further enhancement of the coupling leads into the strong coupling regime with hybridization of cavity and magnon pairs. 
 
For parameters corresponding to $\mathrm{CrI_3}$, the cavity frequency is tuned to be resonant with the edge-magnon pair-creation processes, $\omega_c=2\omega_{K,e}\approx4.5\,\text{THz}$. In order to spectrally resolve the edge-mode contribution, the linewidth of the cavity needs to be smaller than the linewidth of the edge-mode pair or, equivalently in our regime, the linear magnon dissipation rate. For linear magnon dissipation rate $\zeta=0.02J$, the cavity linewidth can be $\kappa<0.02J$. Here, we assume the cavity linewidth to be $\kappa=0.01J\approx0.005\text{THz}$, corresponding to a cavity quality factor $Q= 900$. Further, a cavity with a free spectral range larger than the bandwidth of the material is needed, such that a single cavity mode interacts with the magnon spectrum. This leads to the requirement $\Delta_{FSR}>4.5J\sim2.25\,\text{THz}$ with estimated cavity finesse $F\sim450$. 

In the presence of a cavity the coupling strength per photon is enhanced by the confinement of the EM field inside the cavity. The electric field per photon scales as $\sqrt{1/V_{cav}}$~\cite{Optomagnonics} and consequently, the cavity-magnon coupling per photon reads $\mathcal{G}=\frac{1}{\hbar}\sqrt{\frac{\hbar\omega_c}{2 \epsilon_0 V_{cav}}}\boldsymbol{\Lambda}_{k,s,s'}.\boldsymbol{\epsilon}$, with $\boldsymbol{\Lambda}=\{\boldsymbol{\gamma},\boldsymbol{\mu},\boldsymbol{\nu}\}$ giving the strength of the different photon-magnon interaction processes as in the free-space case. In addition, for a driven cavity, the total intracavity field intensity is determined by the steady-state number of photons circulating in the cavity, $n_{cav}$. This leads to a total cavity-magnon coupling  $\tilde{\mathcal{G}}=\sqrt{n_{cav}}\mathcal{G}$. In our example, the coupling strength between the cavity and edge-magnon pairs has been assumed to be $\tilde{\mathcal{G}}^E_{\nu,K,e,e}=0.05J>\zeta$, a value large enough to obtain amplification. This magnon-cavity coupling is comparable to the effective magnon-drive coupling for pair of magnon edge modes in free space $\tilde{G}^{E}_{\nu,K,e,e}$. 

To estimate the required drive power, we consider a cavity driven through an input port with photon flux $n_{in}=|\langle\hat{c}_{in}\rangle|^2=10^{15}/s $. For $\omega_c=4.5\,\text{THz}$, the corresponding input power can be estimated as $P_{in}=\hbar\omega_c\langle\hat{c}_{in}^\dagger\hat{c}_{in}\rangle \approx3\,\mu\text{W}$. From this, the intracavity photon number can be obtained as $n_{cav}=P_{in}/(\hbar\omega_c\kappa)$, giving $n_{cav}\sim10^5$ for the parameters used here. Assuming a cavity volume $V_{cav}\sim10^{-17}-10^{-16}\,\text{m}^3$, the electric field inside the cavity becomes comparable to the field strength required for parametric amplification of edge modes in free-space. Compared to the free-space pump-probe scheme, where incident intensities of order $10^{12}-10^{14}\,\text{W/m}^2$ are required to achieve amplification, the cavity architecture enables comparable coupling strengths and detection of edge-modes at lower input powers.

\section{\label{sec:Outlook and Conclusion}Outlook and Conclusion}
This work demonstrates that edge modes in magnetic systems can be selectively amplified by external laser driving resonant at twice the magnon frequency. The amplification arises from resonant coupling to two-magnon processes, which selectively populates edge-localized states while only weakly affecting the bulk modes. The system achieves a steady state through a balance between coherent driving and dissipation, with the population stabilized by both linear and nonlinear dissipation channels, ensuring controlled magnon amplification without destabilization. Such tunable steady-state behavior offers a promising route for controlling magnonic edge excitations in topological magnetic materials.

We showed that pump-probe spectroscopy in free-space can directly probe the amplified edge modes. Above a threshold pump intensity, the amplified edge mode generate distinct signatures which can be detected using a probe beam. We showed that the threshold driving intensity depends on the polarization of the incident light. Left circularly polarized incident light requires lower driving intensity to achieve amplification as compared to linearly polarized or right circularly polarized light.

Stability of the bulk magnetization is ensured as long as the magnon-pair frequency is well separated from the bulk band. It should be noted that the bulk-edge separation and estimated values of intensities are highly sensitive to boundary conditions. While a bearded boundary condition results in edge modes well separated from bulk, zigzag boundary condition shifts the edge modes closer to the lower bulk band, making edge-mode resolution difficult and leading to subsequent demagnetization. We showed that an applied static magnetic field can be used to spectrally resolve the signatures from the bulk and the edge states. The linewidth of the pump should be large enough such that the driving field is insensitive to the edge-mode splitting due to the coupling to the light.

To assess the feasibility of observing magnon edge modes in realistic settings, we estimated experimental parameters using monolayer $\mathrm{CrI_3}$ as a prototype representative material, which is known to be a FM TMI.  We estimated a threshold drive intensity required of the order of $10^{13}\,\text{W}/\text{m}^2$ for a linearly polarized pump, operating at a resonant frequency of $4.5\,\text{THz}$ for $0.1\,\text{ns}$, and $10^{11}\, \text{W}/\text{m}^2$ for a left circularly polarized light. Achieving such high average intensities with long pulse duration in the THz gap remains experimentally challenging. However it should be noted that the protocol proposed here is not specific to monolayer $\mathrm{CrI_3}$, and is applicable to monolayer honeycomb ferromagnetic topological insulators e.g. $\mathrm{CrBr_3}$, $\mathrm{CrXTe_3 (X=Si,Ge)}$. Materials with favorable parameters, such as large exchange interaction and low magnon dissipation can enhance the efficiency of amplification, requiring substantially lower driving intensities and shorter timescales for the onset of amplification, making experimental realizations more accessible.

Complementary to pump-probe approaches, coupling the system to THz cavities provides an alternative route for selectively probing and controlling edge modes. Due to the combination of small confinement volume and the photon buildup in the cavity, this platform is suitable to achieve stronger coupling between the magnon and photon as compared to free space. The threshold driving powers required to achieve magnon amplification via pumping the cavity can therefore be considerably lower with respect to the driving intensities required for free space pump probe setups. A cavity with large free spectral range is essential to avoid multiple cavity resonances exciting bulk magnons, enabling selective coupling to edge modes in the desired frequency range.

Recent advances have enabled sub-wavelength confinement of cavity modes in the THz regime~\cite{Faist_2024, Thz}, which could be suitable platforms for our proposal. These platforms support narrow-band operation and can be tuned to match the edge-mode resonance conditions. In particular, THz cavities operating near 4-10 THz with bandwidths of $\sim10\,\text{GHz}$ could provide adequate spectral resolution to resolve edge-mode signatures while avoiding unwanted coupling to bulk modes. Other suitable platforms include THz plasmonic cavities \cite{THz_antennas, Thz_plasmon, Thz_Koppens}, coplanar waveguides \cite{Kipp2024ThzCavity}, and two-dimensional cavity resonators with integrated grating filters \cite{Cavity_Grating}. 
Heterostructures with strong excitonic resonances can also be used to enhance absorption of incident radiation \cite{Kin_Fai_Mak}. 

For a cavity mode at 4.5 THz, with micron scale mode volumes, driven with a few microwatts power, we estimate an intracavity electric field sufficient for edge mode signatures. For $\mathrm{CrI_3}$, therefore, we anticipate that cavity-based probes will operate in the weak-coupling regime, where reflection and transmission spectra remain sensitive to edge-mode resonance features, but without full mode hybridization. Although current cavity architectures do not yet support strong coupling to magnon edge modes, the rapid progress in this field suggests such capabilities may be within reach in the near future. Moreover, other candidates for TMIs can be explored which may exhibit larger spin-dependent electric dipole moment.

Beyond detection, such hybrid platforms could also serve as a medium for long-distance chiral coupling between spin qubits \cite{Mook_Spin-qubit}, thereby offering promising route toward scalable quantum information architectures based on topological magnonics.

\section{\label{Methods}Methods}
\subsection{\label{Methods_2} Pump Probe Spectroscopy}
To probe the effects of the pump-induced excitation, a weak probe beam is introduced. The probe beam interacts with the system, and its response can be measured as changes in the absorbance. The absorbance can be calculated via the susceptibility tensor of the system, $\chi_{\mu \nu}(\omega)$, which can be obtained from the linear response around the non-equilibirum steady state, in terms of the retarded correlation function of the dipole moment~\cite{Kubo, Mook_2024}
\begin{align}
    \chi_{\mu \nu}(\omega)&=-\frac{1}{\hbar} C^R_{P_\mu P_{\nu}}(\omega)\\ \nonumber
    &=i\int_0^{\infty}dt e^{i\omega t}\langle [ P_{\mu}(t),P_{\nu}(0)]\rangle\,,
\end{align}
where $C^R_{P_\mu P_{\nu}}(\omega)$ is the Fourier transform of the retarded correlation function $C^R_{P_\mu P_{\nu}}(t)=-i\theta(t)\langle [ P_{\mu}(t),P_{\nu}(0)]\rangle$ and we set $\hbar=1$. The retarded correlation function can be calculated for the one-magnon and two-magnon contributions to the electric dipole moment. Here, we consider only one diagonal component of the susceptibility, $\chi_{yy}(\omega)$. For one-magnon processes, the correlation can be obtained as 
\begin{align}
    C^{R(1)}_{P_y P_y}(\omega)&=i\int_0^{\infty}dt e^{i\omega t}\langle[ P_{k=0}^{(1)}(t),P_{k=0}^{(1)}(0)]\rangle\\ \nonumber
    &=\sum_{s}\frac{\tilde{G}_{\pi,0,s}^E\tilde{G}_{\pi,0,s}^{E*}}{\omega-\omega_{0,s}+i\zeta^+}\,.
\end{align}
We consider the case in which we probe the system after it has reached its steady state. The expectation value is therefore taken with respect to this steady state with a finite magnon population. However, the one-magnon absorption does not depend on the pumped population. The susceptibility can be obtained as 
\begin{align}
    \text{Im} \chi_{yy}^{(1)}(\omega)=\sum_{s}|\tilde{G}^E_{\pi,0,s}|^2\delta({\omega-\omega_{0,s}})\,.
\end{align}
Now we consider the two-magnon contribution to the correlation function,
\begin{align}
    C^{R(2)}_{P_y P_y}(\omega)&=i\int_0^{\infty}dt e^{i\omega t}\langle[ P_{k}^{(2)}(t),P_{k}^{(2)}(0)]\rangle\,,
\end{align}
using which the two-magnon contribution to the susceptibility can be calculated as
\begin{align}
   \text{Im}\chi_{yy}^{(2)}(\omega)&=\sum_{k,s,s'}|\tilde{G}^E_{\mu,k,s,s'}|^2(n(\omega_{k,s})-n_B(\omega_{k,s'}))\delta({\omega+\omega_{k,s}-\omega_{k,s'}})\\ \nonumber
    &+\sum_{k,s,s'}|\tilde{G}^E_{\nu,k,s,s'}|^2(1+n(\omega_{k,s})+n_B(\omega_{-k,s'}))\delta({\omega-\omega_{k,s}-\omega_{-k,s'}})\\ \nonumber
    &-\sum_{k,s,s'}|\tilde{G}^E_{\nu,k,s,s'}|^2(1+n(\omega_{k,s})+n_B(\omega_{-k,s'}))\delta({\omega+\omega_{k,s}+\omega_{-k,s'}})\,.
\end{align}
Here, we can see the two-magnon contribution to the susceptibility depends on the $n(\omega_{k,s})$, which is the steady state magnon population in magnon mode $\alpha_{k,s}$ due to the pump. In the main text, we plot $\text{Im}\chi_{yy}^{pumped}-\text{Im}\chi_{yy}^{unpumped}$.

\subsection{\label{Methods_3}Input Output formalism}
Here we present the coherent cavity-magnon coupled dynamics. In order to make the pair-creation terms time-independent, we work in a rotating frame where cavity rotates with the frequency of the drive $\omega_d$ and the magnons rotate at $\omega_d/2$.  The equations of motion for the expectation values of the cavity-photon operator and magnon operators can be written, neglecting correlations (Gaussian approximation), as
\begin{equation}
\begin{aligned}
    &\frac{\partial \langle\hat{c}\rangle}{\partial t}=-i(\Delta_{c}-\frac{i}{2}\kappa) \langle\hat{c}\rangle+i\sum_{k,s,s'}\tilde{\mathcal{G}}_{\nu,k,s,s'}^E  \langle\hat{\alpha}_{k,s}\rangle\langle\hat{\alpha}_{-k,s'}\rangle+i(\tilde{G}^E_{\gamma,0,s}-\tilde{\mathcal{G}}^M_{\pi,0,s} )   \langle \hat{\alpha}_{0,s} \rangle e^{i\omega_dt/2} +\sqrt{\kappa_{ex}}\langle\hat{c}_{in}\rangle+\sqrt{\kappa_0} \langle\hat{d}\rangle \\ 
    &\frac{\partial \langle \hat{\alpha}_{k,s}\rangle}{\partial t}=\left(-i\Delta_{k,s}-\frac{\zeta}{2}-\frac{\eta}{2}|\langle\hat{\alpha}_{-k,s}\rangle|^2\right)\langle\hat{\alpha}_{k,s}\rangle+i\sum_{s'} \tilde{\mathcal{G}}_{\nu,k,s,s'}^E  \langle\hat{\alpha}_{-k,s'}^{\dagger}\rangle  \langle \hat{c} \rangle +i(\tilde{\mathcal{G}}^E_{\gamma,0,s}+\tilde{\mathcal{G}}^M_{\pi,0,s})   \langle \hat{c} \rangle e^{-i\omega_dt/2}
    \delta_{k,0}\\ 
    &\frac{\partial \langle \hat{\alpha}_{-k,s}^{\dagger}\rangle}{\partial t}=\left(i\Delta_{k,s}-\frac{\zeta}{2}-\frac{\eta}{2}|\langle\hat{\alpha}_{k,s}\rangle|^2\right)\langle\hat{\alpha}_{-k,s}^{\dagger}\rangle-i\sum_{s'} \tilde{\mathcal{G}}_{\nu,k,s,s'}^E   \langle \hat{c}^\dagger \rangle \langle\hat{\alpha}_{k,s'}\rangle -i(\tilde{\mathcal{G}}^E_{\gamma,0,s}-\tilde{\mathcal{G}}^M_{\pi,0,s} )   \langle \hat{c}^\dagger \rangle e^{i\omega_dt/2} \delta_{k,0}\,,
\end{aligned}
\label{eq:cav}
\end{equation}
where $\hat{c}$ is the cavity photon operator driven by a laser with frequency  $\omega_d$ which is detuned from the cavity frequency by $\Delta_c = \omega_c-\omega_d$. The detuning between the magnon modes and the drive is given by $\Delta_{k,s}=\omega_{k,s}-\omega_{d}/2$. Here $\kappa$ is the total cavity loss, $\kappa_{ex}$ corresponds to loss of cavity to the external port, and $\kappa_0$ denotes the intrinsic losses, such that $\kappa=\kappa_{ex}+\kappa_0$. $\hat{c}_{in}$ is the input field into the cavity and $\hat{d}$ denotes the noise operator associated with intrinsic losses, with $\langle \hat{d}\rangle=0$. The couplings $\tilde{\mathcal{G}}$ correspond to magnon-photon couplings modified due to the confinement of the cavity mode to small volumes~\cite{Optomagnonics}. In our numerical study we assumed the mode overlap between the cavity and magnons to be same for both edge and bulk modes.
We use the input-output formalism~\cite{IO_theory} to obtain the cavity spectrum, according to which $\hat{c}_{out}=\hat{c}_{in}-\sqrt{\kappa_{ex} }\hat{c}$. Using this, we can obtain the reflectance  
\begin{equation}
    r=\frac{\langle \hat{c}_{out}\rangle}{\langle \hat{c}_{in}\rangle}=1-\sqrt{ \kappa_{ex}} \frac{\langle \hat{c}\rangle}{\langle \hat{c}_{in} \rangle}.
\end{equation}

\begin{acknowledgments} 
We thank Sanchar Sharma for the helpful discussions during the initial phase of the project. We would like to thank Tobia Nova, Hans Huebl, Davide Bossini for insights on experimental feasibility. We acknowledge the support from Deutsche Forschungsgemeinschaft (DFG) through SFB TRR 306 'QuCoLiMa' under grant number 429529648, Research Training Group 1995 'Quantum Many Body Methods in Condensed Matter Physics', 'ChiPs' poject number 541503763. We also acknowledge the funding from Bundesministerium für Forchung, Technologie und Raumfahrt (BMFTR) project 'QECHQS' under grant number 16KIS1591 and the computational resources provided by RWTH Aachen University under project ID 'rwth1716'. J.K. acknowledges support from the Deutsche Forschungsgemeinschaft (DFG, German Research Foundation) under grants TRR 360 - 492547816, KN1254/1-2, KN1254/2-1 and under Germany’s Excellence Strategy EXC-2111-390814868, as well as the Munich Quantum Valley, which is supported by the Bavarian state government with funds from the High-tech Agenda Bayern Plus. J.K. further acknowledges support from the Imperial-TUM flagship partnership and the Keck foundation.
\end{acknowledgments}
 
\bibliographystyle{unsrtnat}
\bibliography{MyBib}

\appendix

\pagebreak
\widetext
\begin{center}
\textbf{\large Supplementary Material : Signatures of Topological Magnon Edge States in THz Spectroscopy and Cavity Response}
\end{center}
\renewcommand{\theequation}{S\arabic{equation}}
\renewcommand{\thefigure}{S\arabic{figure}}
\setcounter{equation}{0}
\setcounter{figure}{0}
\setcounter{secnumdepth}{3}
\setcounter{page}{1} 
\makeatletter
\renewcommand{\thesection}{\Alph{section}}
\renewcommand{\thepage}{S\arabic{page}}
\makeatother
\section{\label{AppendixA}Topological Magnon Insulator}

Here we present the details for obtaining the band structure of the ferromagnetic topological magnon insulator on a 2D honeycomb lattice on a ribbon geometry, which is infinite in the $x$-direction and finite in the $y$-direction. We consider the spin Hamiltonian as defined in the main text (Eq.~\ref{eq:Heis}), with nearest-neighbor (n.n) ferromagnetic Heisenberg interaction $J$ and next-nearest-neighbor (n.n.n.) DMI $\mathbf{D}$. The direction of the DMI vector is dictated by the symmetries of the lattice~\cite{Moriya}. In the honeycomb lattice, the n.n. DMI vanishes due to inversion symmetry at the bond center. However, the absence of inversion symmetry at the n.n.n. bond allows for a finite DMI in the direction pointing out of the plane, here along the $z$-direction. Since we are interested in the spectrum of the low energy excitations over the ground state, we perform a Holstein-Primakoff (HP) transformation, transforming the spin operators to  localized bosonic operators $\hat{a}_i$ as 
\begin{align}
    \hat{S}_i^+&=\sqrt{2S}\sqrt{1-\frac{\hat{a}_i^\dagger \hat{a}_i}{2S}}\hat{a}_i \,,\nonumber\\
     \hat{S}_i^-&=\sqrt{2S}\hat{a}_i^\dagger\sqrt{1-\frac{\hat{a}_i^\dagger \hat{a}_i}{2S}}\,, \nonumber\\
     \hat{S}_i^z&=S-\hat{a}_i^\dagger \hat{a}_i\,.
\end{align}
Since the honeycomb lattice has two sublattices, we denote them as A and B and the bosonic operators corresponding to the sublattices are denoted as $\hat{a}_i$ and $\hat{b}_i$. For large-spin systems, the leading order in a 1/S expansion of the HP transformation has the dominant contribution. Thus the higher order of the expansion can be neglected under the linear spin wave approximation, which leads to the quadratic Hamiltonian
\begin{align}
    \hat{H}=zJS\sum_i (\hat{a}_i^\dagger \hat{a}_i+\hat{b}_i^\dagger \hat{b}_i)-JS\sum_{\langle ij \rangle}\hat{a}_i^\dagger \hat{b}_j-DS\sum_{\langle\langle ij \rangle\rangle}e^{i\nu_{ij}\phi}(\hat{a}_i^\dagger \hat{a}_j+\hat{b}_i^\dagger \hat{b}_i)+h.c.\,,
\end{align}
where $z$ is the coordination number of the n.n. sites, $\phi$ is the magnetic flux generated by the DMI (here $\phi=\pi/2$)  and $\nu_{ij}=\pm1$ for clockwise/ anti-clockwise direction of the hopping. 

Since the lattice is infinitely long in the $x$-direction, we use translational invariance to transform into the non-local Fourier basis using
\begin{align}
    \hat{a}_{x,y}^\dagger&=\sqrt{\frac{1}{N_x}}\sum_{k}e^{ikx}\hat{a}_{k,y}^\dagger \nonumber\\
     \hat{a}_{x,y}&=\sqrt{\frac{1}{N_x}}\sum_{k}e^{-ikx}\hat{a}_{k,y}\,,
\end{align}
where $\hat{a}_i(\hat{b}_i)$ have now been written as $\hat{a}_{x,y}(\hat{b}_{x,y})$ to denote the $x$ and $y$ component of the position vector of the atomic sites. The Fourier transformed operator depends on the crystal momenta $k$ along the $x$-direction and the atomic site index $y$ along the $y$-direction.  The Hamiltonian in the transformed basis reads
\begin{align}
    \hat{H}=&zJS\sum_{k,y} (\hat{a}_{k,y}^\dagger \hat{a}_{k,y}+\hat{b}_{k,y}^\dagger \hat{b}_{k,y})-JS\sum_{k,y}(e^{\sqrt{3}ika/2}a_{k,y}^\dagger b_{k,y+1}+e^{-\sqrt{3}ika/2}a^\dagger_{k,y}b_{k,y+1}+a^\dagger_{k,y}b_{k,y-1}) \nonumber\\
    &-DS\sum_{k,y}e^{i\phi}(e^{-\sqrt{3}ika}\hat{a}_{k,y}^\dagger \hat{a}_{k,y}+e^{\sqrt{3}ika/2}\hat{a}_{k,y}^\dagger \hat{a}_{k,y+1}+e^{\sqrt{3}ika/2}\hat{a}_{k,y}^\dagger \hat{a}_{k,y-1}+e^{\sqrt{3}ika}\hat{b}_{k,y}^\dagger \hat{b}_{k,y} \nonumber\\
    &+e^{-\sqrt{3}ika/2}\hat{b}_{k,y}^\dagger \hat{b}_{k,y+1}+e^{-\sqrt{3}ika/2}\hat{b}_{k,y}^\dagger \hat{b}_{k,y-1}) + h.c.\,,
\end{align}
where $a$ is the lattice constant. 

This quadratic Hamiltonian can be written in matrix form by a represented by a $4N_y\times4N_y$ matrix as $\hat{H}=\sum_{k}\hat{\mathbf{A}}_k^\dagger M_k \hat{\mathbf{A}}_k$ in the basis 
\begin{align*}
    \mathbf{\hat{A}_{k}=}&(\hat{a}_{k,1},\hat{a}_{k,2},..,\hat{a}_{k,N},\hat{b}_{k,1},\hat{b}_{k,2},..,\hat{b}_{k,N},\hat{a}_{-k,1}^{\dagger}, 
    \hat{a}_{-k,2}^{\dagger},...,\hat{a}_{-k,N}^{\dagger},\hat{b}_{-k,1}^{\dagger},\hat{b}_{-k,2}^{\dagger},...,\hat{b}_{-k,N}^{\dagger})^{T}\,,
\end{align*}
which can then be diagonalized as $D_k=V^{-1}_{k}M_kV_{k}$ to obtain the quasiparticle energies and wavefunction. For the results in the main text, the matrix $M_k$  was diagonalized numerically to obtain the magnon bands corresponding to the free Hamiltonian $\hat{H}=\sum_{k,s}\hbar\omega_{k,s}\hat{\alpha}_{k,s}^\dagger \hat{\alpha}_{k,s}$, where $\hat{\alpha}_{k,s}$ are the magnon operators with $s$ denoting the $2N_y$ mode-indices for the $N_y$ sites along the $y$-direction, see Fig.~\ref{fig:Main}.

\section{\label{AppendixB}Symmetry Analysis for Spin-dependent electric dipole moment}
In this section, we obtain the spin-dependent electric dipole moment operator allowed by the symmetries of the model. From Eq.~\eqref{eq:Polarization} in the main text, we see that the dipole moment operator can be expanded in terms of the spin operators on the lattice. The dipole moment operator can have a linear spin-dependent term only if the system breaks time-reversal and inversion symmetry. In $\mathrm{CrI_3}$, time-reversal symmetry is broken but the lattice has a center of inversion. The dipole moment is odd under inversion, while the spin operator is an axial operator and thus remains unaffected under inversion. Due to this, the dipole moment operator cannot have a linear spin-dependent contribution.

However, it can depend on quadratic spin-operators on a single site, on pair of spins on different sites, and further higher order terms in spin operators. In this work, we neglect the higher order spin-interactions. We consider a spin-dependent electric dipole moment due to spins on a single site, pair of spins interacting on the n.n. sites, and  the n.n.n. sites. For electric dipole moment induced by spins on a single site
\begin{align}
      \hat{\textbf{P}}_i=\sum_{\mu\nu l}\mathbf{K}_{i\mu\nu}\hat{S}_{i\mu}\hat{S}_{i\nu}\,,
\end{align}
where $i$ is the site index, and $\mu, \nu $ are the components of the spin operators.  The form of the tensor coefficient $K_{i\mu\nu}^{\alpha}$ can be obtained based on the symmetry of the lattice and boundary conditions. We consider a ribbon geometry with open boundaries along $x$-direction and closed boundary along $y$-direction. The symmetries at each site on a honeycomb lattice includes a $C_3$ rotation around the z-axis, $C_2$ rotation around the axis perpendicular to the zigzag/ bearded boundary (i.e., y-axis) and absence of inversion symmetry at the lattice sites. Under a symmetry operator $\hat{R}$, the coefficient matrix transforms as~\cite{Symmetry_2017}
\begin{align}
    R_{\alpha\beta}K_{i\gamma\delta}^{\beta}=R_{\gamma\mu}^TK_{i\mu\nu}^{\alpha}R_{\nu\delta}\,,
    \label{eq:transfrom}
\end{align}
which for the above case reduces to
\begin{align}
   K^{x} & =\left(\begin{array}{ccc}
0 & K_{xy}^{x} & 0 \nonumber\\
K_{xy}^{x} & 0 & K_{yz}^{x} \nonumber\\
0 & K_{yz}^{x} & 0
\end{array}\right)\,,\\
K^{y} & =\left(\begin{array}{ccc}
K_{xy}^{x} & 0 &- K_{yz}^{x} \nonumber\\
0 & -K_{xy}^{x} & 0 \nonumber\\
-K_{xz}^{y} & 0 & 0
\end{array}\right)\,.
\end{align}
The quadratic term of the induced electric dipole moment operator depending on spins on single site is given as
\begin{align}
    \hat{P}_i^x=&K_{xy}^x (\hat{S}_{ix}\hat{S}_{iy}+\hat{S}_{iy}\hat{S}_{ix})+K_{yz}^x (\hat{S}_{iy}\hat{S}_{iz}+\hat{S}_{iz}\hat{S}_{iy})\,,\nonumber\\
     \hat{P}_i^y=&K_{xy}^x (\hat{S}_{ix}^2-\hat{S}_{iy}^2)-K_{yz}^x (\hat{S}_{ix}\hat{S}_{iz}+\hat{S}_{iz}\hat{S}_{ix})\,,\nonumber\\
     \hat{P}_i^z=&0\,.
\end{align}
Similarly, we consider the quadratic spin-dependent electric dipole moment originating from spin interactions on the neighboring sites. The dipole moment operator induced due to a pair of spins can be separated into a symmetric contribution and an anti-symmetric contribution as 
\begin{align}
    \hat{\textbf{P}}_{ij}=&\textbf{A}_{\mu\nu}\hat{S}_{i\mu}\hat{S}_{j\nu}+\textbf{B}_{\mu\nu}(\hat{S}_{i\mu}\hat{S}_{i\nu}-\hat{S}_{j\mu}\hat{S}_{j\nu}) \nonumber+\textbf{C}_{\mu\nu}\hat{S}_{i\mu}\hat{S}_{j\nu}+\textbf{D}_{\mu\nu}(\hat{S}_{i\mu}\hat{S}_{i\nu}+\hat{S}_{j\mu}\hat{S}_{j\nu})\,,
    \label{eqn:spin_expand}
\end{align}
where $\textbf{C}_{\mu\nu}=-\textbf{C}_{\nu\mu}$ and the rest of the coefficients are symmetric under exchange of $\mu, \nu$. The coefficients which are symmetric under exchange of $\mu,\nu$ transform as Eq.~\ref{eq:transfrom} under rotations. For an antisymmetric tensor, there is an additional factor taking care of spin-exchange as
\begin{align}
    R_{\alpha\beta}K_{\gamma\delta}^{\beta}=\eta_{l\xrightarrow{}m}R_{\gamma\mu}^TK_{\mu\nu}^{\alpha}R_{\nu\delta}\,,
\end{align}
where $\eta_{i\xrightarrow{}j}=-1$ if there is an interchange of spins between sites $l$ and $m$ due to the rotation and $\eta_{i\xrightarrow{}j}=1$ otherwise.

For the zigzag and bearded boundary condition, the symmetry at the center of the n.n. bond aligned along the $y$-direction is $C_{2h}\equiv\{E,i,C_{2y},\sigma_y\}$. This leads to the absence of symmetric spin contributions from n.n. bonds bonds. So for nearest neighbor bonds the effective dipole moment has the form
\begin{align}
    \hat{\textbf{P}}_{<ij>}=\textbf{B}_{\mu\nu}(\hat{S}_{i\mu}\hat{S}_{i\nu}-\hat{S}_{j\mu}\hat{S}_{j\nu}) +\textbf{C}_{\mu\nu}\hat{S}_{i\mu}\hat{S}_{j\nu}\,.
\end{align}
Due to the antisymmetry $\textbf{C}_{\mu\nu}=-\textbf{C}_{\nu\mu}$, the second terms can be rewritten as $\textbf{C}_{\mu\nu}(\hat{S}_{i\mu}\hat{S}_{j\nu}-\hat{S}_{i\nu}\hat{S}_{j\mu})$. This can be expressed equivalently as $\textbf{C}_{\lambda}(\hat{\textbf{S}}_{i}\times\hat{\textbf{S}}_{j})_\lambda$, where $\textbf{C}_{\lambda}=\epsilon_{\lambda\mu\nu}\textbf{C}_{\mu\nu}$, with $\epsilon_{\lambda\mu\nu}$ being the Levi-Civita tensor. Rewriting in this form reduces the number of symmetry allowed independent coefficient matrix components. Under the symmetry operations at the bond center, the only non-zero contributions to coefficient matrix $\textbf{C}_\lambda$ are $C_x^x$, $C_x^z$, $C_y^y$, $C_z^x$, and $C_z^z$, where the upper index of the matrix element denotes the direction of the vector coefficient matrix $\textbf{C}$  and the lower index denotes the component $\lambda$ of the vector $(\hat{\textbf{S}}_{i}\times\hat{\textbf{S}}_{j})$.
The coefficient matrix $\textbf{B}_{\mu\nu}$ reduce to 
\begin{align}
    B^{x}=\left(\begin{array}{ccc}
0 & B_{xy}^{x} & 0\\
B_{yx}^{x} & 0 & B_{yz}^{x}\\
0 & B_{zy}^{x} & 0
\end{array}\right);B^{y}=\left(\begin{array}{ccc}
B_{xx}^{y} & 0 & B_{xz}^{y}\\
0 & B_{yy}^{y} & 0\\
B_{zx}^{y} & 0 & B_{zz}^{y}
\end{array}\right);B^{z}=\left(\begin{array}{ccc}
0 & B_{xy}^{z} & 0\\
B_{yx}^{z} & 0 & B_{yz}^{z}\\
0 & B_{zy}^{z} & 0
\end{array}\right)\,.
\end{align}
Using the above symmetry-allowed matrix coefficients, the electric dipole moment induced on the n.n. bond is given as
\begin{align}
P_{<ij>}^{u}&=C_{u}^{u}(S_{i}\times S_{j})_{u}+B_{vv}^{u}(S_{iv}^{2}-S_{jv}^{2})+B_{uu}^{v}(S_{iu}^{2}-S_{ju}^{2})+B_{ww}^{v}(S_{iw}^{2}-S_{jz}^{2})\\ \nonumber
	&+B_{vw}^{u}(S_{iv}S_{iw}+S_{iw}S_{iv}-S_{jv}S_{jw}-S_{jw}S_{jv})\\ \nonumber
    P_{<ij>}^{v}&=C_{v}^{v}(S_{i}\times S_{j})_{v}+C_{w}^{v}(S_{i}\times S_{j})_{w}+B_{vu}^{v}(S_{iv}S_{iu}+S_{iu}S_{iv}-S_{jv}S_{ju}-S_{ju}S_{jv})\\ \nonumber
	&+B_{uw}^{u}(S_{iu}S_{iw}+S_{iw}S_{iu}-S_{ju}S_{jw}-S_{jw}S_{ju})\\ \nonumber
P_{<ij>}^{w}&=C_{v}^{w}(S_{i}\times S_{j})_{u}+C_{w}^{w}(S_{i}\times S_{j})_{w}+B_{vu}^{w}(S_{iv}S_{iu}+S_{iu}S_{iv}-S_{jv}S_{ju}-S_{ju}S_{jv})\\ \nonumber
	&+B_{uw}^{w}(S_{iu}S_{iw}+S_{iw}S_{iu}-S_{ju}S_{jw}-S_{jw}S_{ju})\,,
\end{align}
which is expressed in a transformed basis $\{u,v,w\}$ where, $u$ is the direction along the bond, $v$ is perpendicular to the bond in the plane and $w$ is perpendicular to the plane. Here we see antisymmetric contributions of the form $(\hat{S}_{iu}^2-\hat{S}_{ju}^2)\hat{u}$, $ \Big(\hat{\textbf{S}}_{i}\times\hat{\textbf{S}}_{j}\Big)\times\hat{u}$ where $\hat{u}$ is the bond-direction between the n.n. sites.
A n.n.n. bond aligned along the $x$-direction only has $\sigma_x$ mirror symmetry, which allows for both symmetric and anti-symmetric contributions in the spins. The matrix coefficients in this case take the form
\begin{align}
    C=\left(\begin{array}{ccc}
C_{x}^{x} & 0 & 0\\
0 & C_{y}^{y} & C_{z}^{y}\\
0 & C_{y}^{z} & C_{z}^{z}
\end{array}\right)\,,
\end{align}
\begin{align}
    B^{x}=\left(\begin{array}{ccc}
B_{xx}^{x} & 0 & 0\\
0 & B_{yy}^{x} & B_{yz}^{x}\\
0 & B_{yz}^{x} & B_{zz}^{x}
\end{array}\right);\left(\begin{array}{ccc}
0 & B_{xy}^{y} & B_{xz}^{y}\\
B_{xy}^{y} & 0 & 0\\
B_{xz}^{y} & 0 & 0
\end{array}\right);\left(\begin{array}{ccc}
0 & B_{xy}^{z} & B_{xz}^{z}\\
B_{xy}^{z} & 0 & 0\\
B_{xz}^{z} & 0 & 0
\end{array}\right)\,.
\end{align}
The coefficient matrices $\textbf{A}_{\mu\nu}$ and $\textbf{D}_{\mu\nu}$ have the same symmetry allowed form as $\textbf{B}_{\mu\nu}$. With this, the electric dipole moment on the n.n.n. bond can be expressed as
\begin{align}
    P_{<<ij>>}^{u}&=C_{u}^{u}(S_{i}\times S_{j})_{u}+B_{uu}^{u}(S_{iu}^{2}-S_{ju}^{2})+B_{vv}^{u}(S_{iv}^{2}-S_{jv}^{2})+B_{ww}^{u}(S_{iz}^{2}-S_{jz}^{2})\\ \nonumber
	&+B_{vw}^{u}(S_{iv}S_{iw}+S_{iw}S_{iv}-S_{jv}S_{jw}-S_{jw}S_{jv})+D_{uu}^{u}(S_{iu}^{2}+S_{ju}^{2})+D_{vv}^{u}(S_{iv}^{2}+S_{jv}^{2})+D_{ww}^{u}(S_{iz}^{2}+S_{jz}^{2})\\ \nonumber
	&+D_{vw}^{u}(S_{iv}S_{iw}+S_{iw}S_{iv}+S_{jv}S_{jw}+S_{jw}S_{jv})+2A_{uu}^{u}S_{iu}S_{ju}+2A_{vv}^{u}S_{iv}S_{jv}+2A_{ww}^{u}S_{iz}S_{jz}\\ \nonumber
	&+A_{vw}^{u}(S_{iv}S_{jw}+S_{iw}S_{jv})\\ \nonumber
P_{<<ij>>}^{v}&=C_{v}^{v}(S_{i}\times S_{j})_{v}+C_{w}^{v}(S_{i}\times S_{j})_{w}+B_{uv}^{v}(S_{iu}S_{iv}+S_{iv}S_{iu}-S_{ju}S_{jv}-S_{jv}S_{ju})\\ \nonumber
	&+B_{uw}^{v}(S_{iu}S_{iw}+S_{iw}S_{iu}-S_{ju}S_{jw}-S_{jw}S_{ju})+D_{uv}^{v}(S_{iu}S_{iv}+S_{iv}S_{iu}+S_{ju}S_{jv}+S_{jv}S_{ju})\\ \nonumber
	&+D_{uw}^{v}(S_{iu}S_{iw}+S_{iw}S_{iu}+S_{ju}S_{jw}+S_{jw}S_{ju})+A_{uv}^{v}(S_{iu}S_{jv}+S_{iv}S_{ju})+A_{uw}^{v}(S_{iu}S_{jw}+S_{iw}S_{ju})\\ \nonumber
P_{<<ij>>}^{w}&=C_{v}^{w}(S_{i}\times S_{j})_{v}+C_{w}^{w}(S_{i}\times S_{j})_{w}+B_{uv}^{w}(S_{iu}S_{iv}+S_{iv}S_{iu}-S_{ju}S_{jv}-S_{jv}S_{ju})\\ \nonumber
	&+B_{uw}^{w}(S_{iu}S_{iw}+S_{iw}S_{iu}-S_{ju}S_{jw}-S_{jw}S_{ju})+D_{uv}^{w}(S_{iu}S_{iv}+S_{iv}S_{iu}+S_{ju}S_{jv}+S_{jv}S_{ju})\\ \nonumber
	&+D_{uw}^{w}(S_{iu}S_{iw}+S_{iw}S_{iu}+S_{ju}S_{jw}+S_{jw}S_{ju})+A_{uv}^{w}(S_{iu}S_{jv}+S_{iv}S_{ju})+A_{uw}^{w}(S_{iu}S_{jw}+S_{iw}S_{ju})\,.
\end{align}
Here we see additional spin-interaction contributions of the form $(\hat{\textbf{S}}_i \cdot\hat{\textbf{S}}_j)\hat{\textbf{u}}$ and $(\hat{S}_{iu}^2+\hat{S}_{ju}^2)\hat{\textbf{u}}$, which are absent on the n.n. bonds. 

\section{\label{AppendixC}Interaction Hamiltonian}
In this section, we formulate the effective magnon-photon coupling induced by the THz electromagnetic (EM) field. The interaction Hamiltonian in the discretized form is given as 
\begin{align}
    \hat{H}_{int}=-\sum_i \hat{\mathbf{P}}_i.\hat{\mathbf{E}}-\sum_i\hat{\mathbf{M}}_i.\hat{\mathbf{B}}\,,
\end{align}
where the summation is over the atomic sites of the semi-infinite TMI strip.

In order to study the coupling between  magnons and the THz field, we can map the spins on the lattice to the second quantized bosonic form using a HP transformation up to leading order and subsequent Fourier transform as described in~SM~\ref{AppendixA}. In the magnon basis, the total magnetic moment $\hat{\mathbf{M}}=\sum_i g\mu_B \hat{\mathbf{S}}_i$ can be expressed as
\begin{align}
    \hat{M}^x=&\sqrt{\frac{NS}{2}}g\mu_B\sum_{s}\pi^{x}_{0,s}\hat{\alpha}_{0,s}+h.c. \,,\nonumber\\
     \hat{M}^y=&\sqrt{\frac{NS}{2}}g\mu_B\sum_{s}\pi^{y}_{0,s}\hat{\alpha}_{0,s}+h.c. \,,\nonumber\\
     \hat{M}^z=&g\mu_BNS-g\mu_B\sum_{k,s}\hat{\alpha}^\dagger_{k,s}\hat{\alpha}_{k,s}\,,
\end{align}
where we have used the unitary transformation $\hat{\alpha}_{k,s}=\sum_y{V}_{k,s,y}\hat{a}_{k,y}$ (see SM~\ref{AppendixA}) implemented numerically, and $\boldsymbol{\pi}_{k,s}=(\pi_{k,s}, i\pi_{k,s})$, where $\pi_{k,s}=\frac{1}{\sqrt{N_y}}\sum_yV^{*}_{k,y,s}$ comes from diagonalizing the system in the finite direction of the lattice (also performed numerically).

\begin{figure}[hbt!]
	\begin{center}
        \includegraphics[width = 1.0 \textwidth]{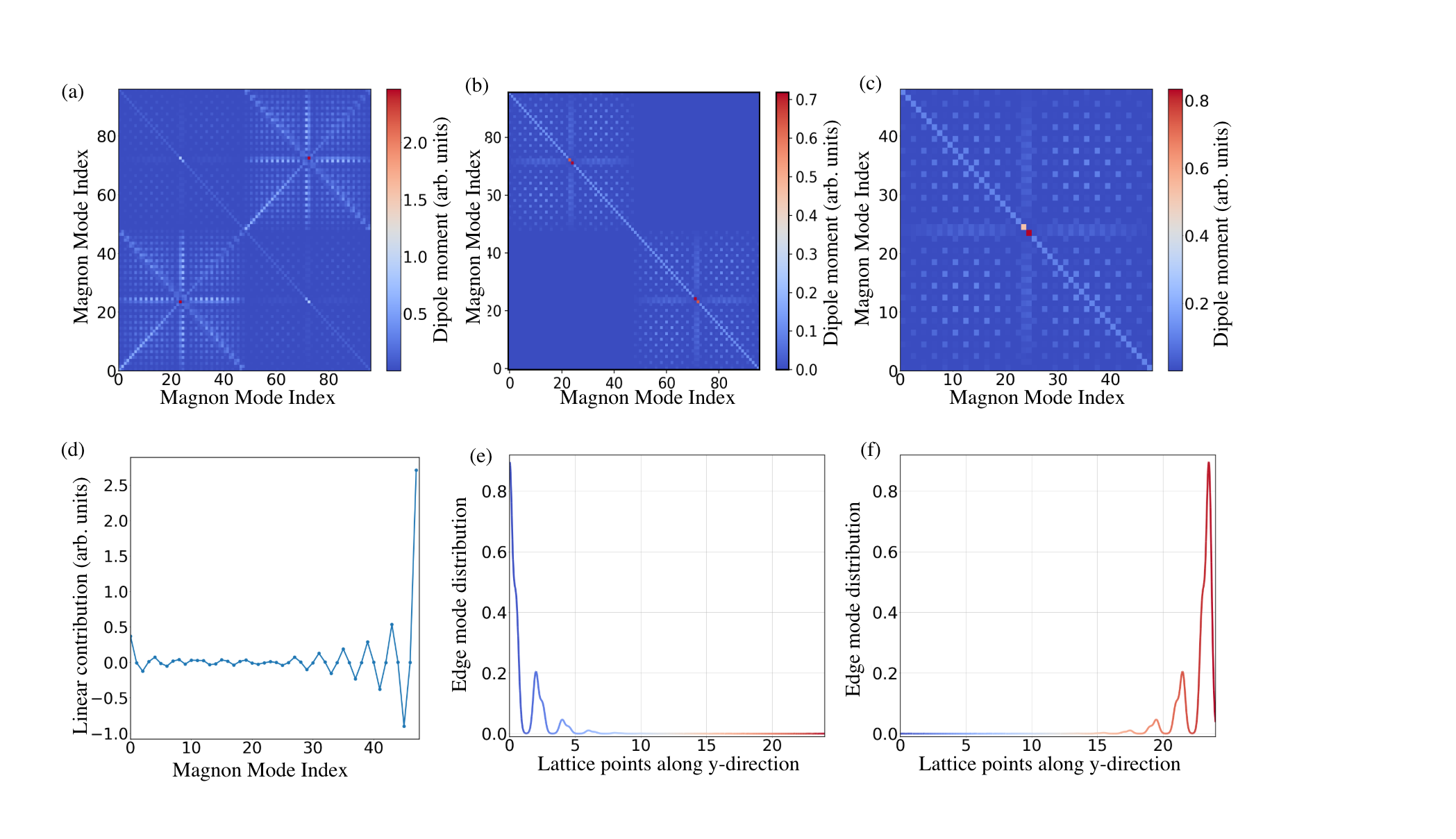}
    \captionsetup{width=0.95\linewidth}
        \captionsetup{justification=raggedright,singlelinecheck=false}
		\caption{\textbf{Magnon-mode contributions to the electric dipole moment at K}  Matrix representation of the magnon-mode contribution to $P_k^{(2)}$ for (a) $y$-polarization and (b) $x$-polarization. The first and fourth quadrants represents scattering between magnons with coefficients $\boldsymbol{\mu}_{K,ss'}$. The second and third quadrants denote two-magnon creation and annihilation denoted by $\boldsymbol{\nu}_{K,ss'}$. (c) Zoom in onto the second quadrant of (a), where the maximum coupling can be seen for magnons with mode index 23 and 24 corresponding to both edges. (d) The one-magnon contribution to electric dipole moment $\pi_{0,s}$ for different magnon mode index. (e,f) Mode-profiles for magnon-mode indexes 23 and 24, showing that these are the modes localized at the edges. }
        \label{fig:Dipole_two_mag}
	\end{center}
\end{figure}

The spin-dependent electric dipole moment can be expressed similarly in terms of magnon operators as (see Eqs.~\ref{eq:5},\ref{eq:6} in the main text)
\begin{align}
    \hat{\mathbf{P}}_{k=0}^{(1)}&=\sum_{s}\boldsymbol{\gamma}_{0,s}\hat{\alpha}_{0,s}+h.c.\\
    \hat{\mathbf{P}}_{k}^{(2)}&=\sum_{s,s'}\boldsymbol{\mu}_{k,ss'}\hat{\alpha}_{k,s}^{\dagger}\hat{\alpha}_{k,s'}+\boldsymbol{\nu}_{k,ss'}\hat{\alpha}_{k,s}\hat{\alpha}_{-k,s'}+h.c.  \,,
\end{align}
where $\hat{\mathbf{P}}_{k=0}^{(1)}$ and $\hat{\mathbf{P}}_{k}^{(2)}$ are the effective electric dipole moment to linear and quadratic order in the magnon operators, respectively. The coefficient $\boldsymbol{\gamma}_{0,s}$ denotes the strength of one-magnon contribution and $\boldsymbol{\mu}_{k,ss'}$, $\boldsymbol{\nu}_{k,ss'}$ denote two-magnon contributions to the electric dipole moment, obtained based on the symmetry-allowed spin expansion of electric dipole moment (see SM~\ref{AppendixB}). It should be stressed that the magnon-pair coupling originates from the n.n.n. spin interactions. The dependence of the vector coefficients on $k$ and $s$ is dictated by the lattice symmetries, boundary conditions, and the effective magnon band-structure on the finite lattice. To examine these dependencies, we numerically obtained the one-magnon and two-magnon contributions to the electric dipole moment for the ribbon geometry taking 24 sites along $y$-direction. As shown in (Fig.~\ref{fig:Dipole_two_mag}), the maximum contribution to the electric dipole moment comes from the edge modes. In the plot, we show the relative contributions of magnon-mode couplings to the electric dipole moment. The magnon-pair contribution can be seen to be dominant for $y$-polarization of the incident light, while magnon-scattering is dominant for $x$-polarization. The coupling strengths can be obtained via a microscopic model, as discussed in SM~\ref{AppendixD}.

\section{\label{AppendixD}Magnon-Photon Interaction strengths}
In the previous section, we derived the spin-dependent electric dipole moment allowed by symmetry which couples the TMI to an external electric field. These terms originate due to charge fluctuations on the magnetic lattice. This can be modeled microscopically starting from a multi-orbital Hubbard-Kanamori model in presence of spin-orbit coupling and considering virtual hopping processes of the electrons between the magnetic sites~\cite{Adrien_d5,microscopics}. A full microscopic derivation of the spin-dependent electric  dipole moment for the FM TMI requires considering virtual hopping processes up to n.n.n. sites, which is essential to obtain the n.n.n. DMI in the effective low energy model. A complete derivation up to n.n.n. is technically involved and out of the scope of this work. In this section we present a simplified model considering virtual hoppings only up to n.n. sites and provide estimates for the effective electric dipole moment within this model. We use a scaling argument to extend the estimates to n.n.n. contributions.

The full Hubbard-Kanamori Hamiltonian can be expressed as~\cite{microscopics} 
\begin{equation}
    H=H_{hop}+H_{CF}+H_K + H_{pd},
\end{equation}
where 
\begin{align}
    &H_{hop}=\sum_{<ij>,\alpha,\beta,\sigma,\sigma'}T_{ij}c_{i,\alpha,\sigma}^{\dagger}c_{j,\beta,\sigma'}\,,\\
    &H_{CF}=\Delta_{CF}\sum_{i,\alpha\epsilon e_{g},\sigma}c_{i,\alpha,\sigma}^{\dagger}c_{i,\alpha,\sigma}\,,\\
   & H_{K}	=\sum_{i,\sigma,\sigma',\alpha,\beta}Un_{i,\alpha,\sigma}n_{i,\alpha,\sigma'}+\frac{U'}{2}n_{i,\alpha,\sigma}n_{i,\beta,\sigma'} -\frac{J_{H}}{2}c_{i,\alpha,\sigma'}^{\dagger}c_{i,\beta,\sigma}^{\dagger}c_{i,\beta,\sigma'}c_{i,\alpha,\sigma}+J_{H}c_{i,\alpha,\uparrow}^{\dagger}c_{i,\alpha,\downarrow}^{\dagger}c_{i,\beta,\uparrow}c_{i,\beta,\downarrow}\,,\\
   &H_{pd}=\Delta_{pd}\sum_{i,m,\alpha,\sigma,\gamma\epsilon p}c_{i,\alpha,\sigma}^{\dagger}c_{m,\gamma,\sigma}+c_{m,\gamma,\sigma}^{\dagger}c_{i,\alpha,\sigma}\,.
\end{align}  
Here, $T_{ij}$ is the hopping matrix with $c_{i,\alpha,\sigma}^{(\dagger)}$ being the fermionic annihilation (creation) operator on TM ion site $i$ in $d$-orbital $\alpha$ with spin $\sigma$, and m is the site index of ligands with $\gamma$ denoting the $p$-orbitals. $H_{CF}$ is the crystal field splitting energy due to the octahedral arrangement of the ligands around the TM ions, where $\Delta_{CF}$ is the energy cost for electrons in $e_g$ orbitals. $U,U',J_H$ are the intra-orbital and inter-orbital Coulomb interaction, and Hund's coupling, respectively. In addition, electrons can also hop between the TM ions via a superexchange pathway mediated by the ligands. The energy cost to hop from a TM to a ligand is denoted as $\Delta_{pd}$.

The effective low energy spin Hamiltonian up to n.n. interactions can be obtained from perturbation theory as~\cite{Valenti_n.n.n, microscopics}
\begin{equation}
    H=\sum_{<ij>,\gamma}\Big[J\textbf{S}_i\textbf{S}_j+ K^\gamma S_i^\gamma S_j^{\gamma}\Big]\,,
\end{equation}
where $J_{ij}$ and $K^\gamma$ is the Heisenberg and Kitaev interactions, with $\gamma=\{x,y,z\}$ corresponding to bonds along $\{X,Y,Z\}$. While we do not derive n.n.n. spin interactions, it is to be noted that at very high fields, the magnon spectrum of Kitaev model reduces to that of the n.n. Heisenberg model with n.n.n. DMI~\cite{Theory_TMI2}, which justifies our microscopic model in the regime where a strong B-field is applied.  

Similarly, the low energy effective electric dipole moment can also be obtained with the strong coupling expansion. For a two site model, the dipole moment operator can be expressed as $\textbf{P}_{ij}=e\delta n  \textbf{r}_{ij}$, where $e$ is the electron charge and $\delta n$ quantifies the charge fluctuation between the two sites and $\textbf{r}_{ij}=\textbf{r}_j-\textbf{r}_i$ is the distance between the two TM sites. A perturbative expansion of the charge fluctuation on a site considering virtual hopping processes results in an effective polarization in the low-energy subspace~\cite{Adrien_d5}
\begin{align}
    \textbf{P}_{ij}^{eff}&=\left[H^{ij}_{hop}\frac{Q}{(E_0-H_0)^2}H^{ji}_{hop}-H^{ij}_{hop}\frac{Q}{(E_0-H_0)^2}H^{ji}_{hop}\right]e \textbf{r}_{ij}\,,
\end{align}
where we consider a two-site model with ground state electronic configuration $t^3_{2g}-t^3_{2g}$, with $Q$ being the projection onto excited states after one electron hopping i.e., states with electronic occupancies $t_{2g}^2-t_{2g}^4$ or $t_{2g}^2-t_{2g}^3e_g^1$. $E_0$ is the ground state energy and $H_0$ is the on-site Hamiltonian including contributions from $H_K$ and $H_{CF}$ on a single site. From such an effective model, it has been shown that the spin-dependent electric dipole moment can have  dependencies of the form ~$\frac{t^4}{\Delta_{pd}^2 U^2}(S_i\times S_j)$ in the leading order, where $t$ is the hopping strength between n.n. sites. For $\mathrm{CrI}_3$, these values have been estimated as $t=0.7eV$, $\Delta_{pd}=2eV$ and $U=3eV$~\cite{microscopics}. Using these values, $\frac{t^4}{\Delta_{pd}^2 U^2} \approx 6.7 \times 10^{-3}$ and hence the effective charge and dipole moment can be approximated as 
\begin{align}
    q_{eff}=e\delta n \approx e \frac{t^4}{\Delta_{pd}^2 U^2} =1.26\times10^{-21}\,\text{C}\,,\\
     | P_{<ij>}|=q_{eff} a_{0}   \approx 8.9 \times 10^{-31} \,\text{Cm}\,,
     \label{eq:dipole_moment_micro}
\end{align}
where $a_0 =7\,\mathring{\text{A}}$ is the lattice constant of $\mathrm{CrI}_3$. For a typical ribbon geometry with unit cell volume of roughly $900\mathring{\text{A}}^3$ \footnote{Data retrieved from the Materials Project for CrI3 (mp-1213805) from database version v2025.09.25.}, the electric dipole moment density can be estimated to be $\approx 1.01\,m \text{C}/\text{m}^2$. This would correspond to the dominant contributions to $\boldsymbol{\gamma},\boldsymbol{\mu}$, except the $k$ and $s$ dependencies which come from the lattice structure. The effective charge for $\mathrm{CrI}_3, \mathrm{CrBr}_3$ have been estimated in earlier works~\cite{Mook} to be $2 \times 10^{-22} \,\text{C}$, which is close to our estimates of effective charge.  Now for the next-nearest-neighbor interaction, the leading order interaction can be estimated to be suppressed by ~$t_{n.n.n}/U^2$ which for a honeycomb geometry leads to $P_{n.n.n} \approx 0.05 P_{n.n.}=4.45\times10^{-32}\,\text{Cm}$, corresponding to $50\mu \text{C}/\text{m}^2$.  This estimate enters in the coefficients $G^{E(n.n.n.)}_{\nu,k,s,s'}$. If we include the lattice dependencies, we obtain the two-magnon contribution $\boldsymbol{\nu}_{K,e,e}$ along the $y$-direction for the edge mode  $\approx 50\,\mu \text{C}/\text{m}^2$ whereas the bulk contribution is suppressed by a factor of at least $10^{-3}$.

There can be other contributions to the spin-dependent electric dipole moment which arise in the presence of strong spin-orbit coupling or Hund's coupling~\cite{Adrien_d5}. In~\cite{Polarization_spin_2007}, it has been shown that different spin-interaction mechanisms responsible for an induced electric dipole moment can have different strengths and depend significantly on the material parameters. The magnitude of these contributions and the spin-dependence originating from these terms are not explicitly discussed here. However, it can be expected that the magnitude of the electric dipole moment depends on $J_H$  and new spin-interactions which are absent for $J_H\rightarrow 0$ can arise. Systems with large spin-orbit coupling on the ligands can lead to a larger contribution to the p-d hybridization mechanism, consequently leading to a contribution of the form $\hat{S}_{iu}^2-\hat{S}_{ju}^2$. For $\mathrm{CrI}_3$, we can expect these contributions to be significant. In this work, we have included these terms based on a symmetry analysis without explicitly discussing their microscopic origin.

The magnitude of the electric field depends on the intensity of the drive.  For a sample of $\mathrm{CrI}_3$ with dimensions of roughly $25\times 200$ sites, i.e., $2.5 \times 10^{-15}\,\text{m}^2$ area, illuminated by a laser source of power 10\,mW, the intensity is approximately $4 \times10^{12}\,\text{W/m}^2$. This results in $|E_c|= \sqrt{2I/c\epsilon_0} \approx 5.5 \times 10^7 \,\text{V/m} $.  This is roughly of the same order of magnitude of a pulsed laser field of fluence $\approx 10\,\text{mJ/cm}^2$ for 0.6\,ps. The magnitude of the magnetic field is then $|E_c|/c \approx 0.18\,\text{T}$.

Using the above estimates for the spin-dependent electric dipole moment and the EM field strengths, we can obtain the interaction strength between the TMI and the EM field, $\tilde{G}_{\Lambda,k,s,s'}^{E(n.n.)}=\frac{1}{\hbar}\sqrt{\frac{2I}{c\epsilon_0}}\boldsymbol{\Lambda}^{n.n.}_{k,s,s'}.\boldsymbol{\epsilon} \sim 0.31\,\text{meV}$ which in units of $J$ is approximately  $0.15J$ for the coupling due to n.n. spin-interactions. Similarly, n.n.n. spin interactions lead to a coupling strength of $0.015\text{meV}$, equivalent to $0.0075J$ . The coupling originating from n.n.n. interactions is important because this gives rise to the parametric processes. For parametric amplification for magnons  with linear magnon dissipation $\zeta=0.04J$, the magnon-drive coupling required is $\tilde{G}_{\Lambda,k,s,s'}^{E(n.n.n.)}>0.04J$. The intensity of the drive required to achieve this coupling can be estimated to be $2.98\times 10^{13}\,\text{W/m}^2$ corresponding to an electric field of $E_c=1.7\times10^{8}\,V/m$. In the main text and in other figures of SM, we vary the pump intensity from $1.86\times 10^{11}\text{W/m}^2$ to $1.16\times 10^{14}\,\text{W/m}^2$. This is same as varying the coupling strength $\tilde{G}_{\nu,k,s,s'}^{E}$ which originates from the n.n.n. spin interactions up to $0.25J$ in steps of $0.01J$.

\section{\label{AppendixE}Edge-Mode Splitting}
\begin{figure}[hbt!]
	\begin{center}
        \includegraphics[width = 0.55 \textwidth]{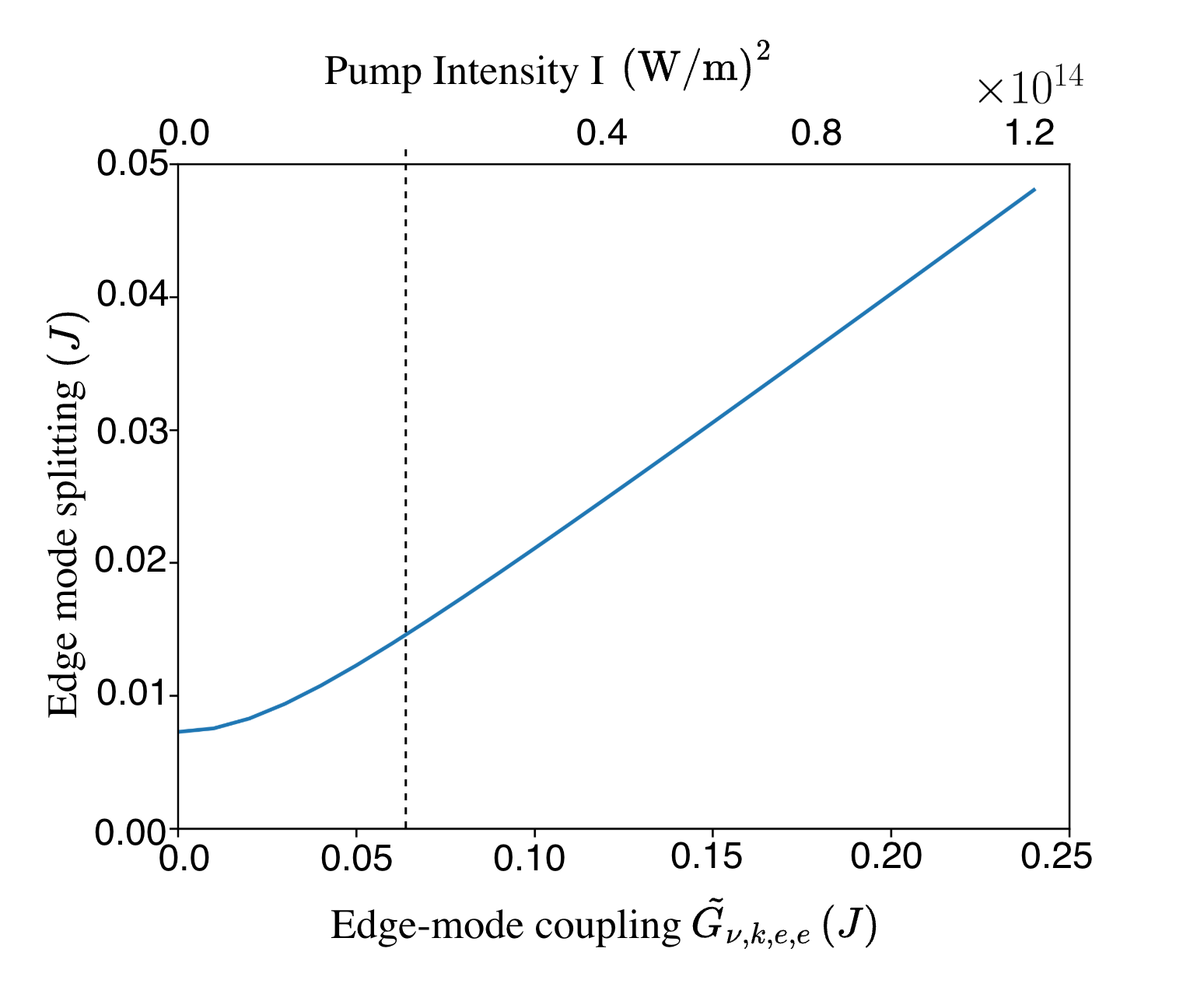}
        \captionsetup{width=0.95\linewidth}
        \captionsetup{justification=raggedright,singlelinecheck=false}
		\caption{\textbf{Edge mode splitting} The edge-mode splitting increases as a function of the driving pump intensity. The bottom $x$-axis, shows the edge mode coupling strength $\tilde{G}_{\nu,K,e,e}$ varying between $0.01-0.25J$ and the upper axis shows the corresponding driving intensity. The threshold drive intensity is denoted by the dashed line.}
        \label{fig:splitting}
	\end{center}
\end{figure}

The coupling to the external electromagnetic field acts as a perturbation that lifts the degeneracy of the edge modes at the $K$ point. To quantify this effect, we diagonalize the Hamiltonian $\hat{H}=\hat{H}_{FM}+\hat{H}_{int}$ numerically by changing the intensity $I$ of the driving field. The classical drive amplitude c scales with $\sqrt{I}$ and consequently the magnon-drive coupling also scales with the intensity of the drive as $\sqrt{I}$. So, varying the intensity results in an increase in the strength of the perturbation, resulting in the splitting of the edge modes as depicted in Fig.~\ref{fig:splitting}. 

The drive-induced splitting limits the intensity and the minimum linewidth of the incident pump laser. The linewidth of the pump should be larger than the edge mode splitting in order to simultaneously drive both edge modes participating in the parametric pair-creation process near the degeneracy point $K$. For the range of drive intensities considered in this work, the splitting between the edge modes can go up to $0.05J$. Using the energy scale $J=0.5\,\mathrm{THz}$, the edge-mode splitting corresponds to $0.025\,\text{THz}$. This sets a lower bound on the required pump linewidth.
\section{\label{Appendix_Demag}Demagnetization effects}
\begin{figure}[hbt!]
	\begin{center}
        \includegraphics[width = 0.92 \textwidth, trim={0 500 0 0},clip]{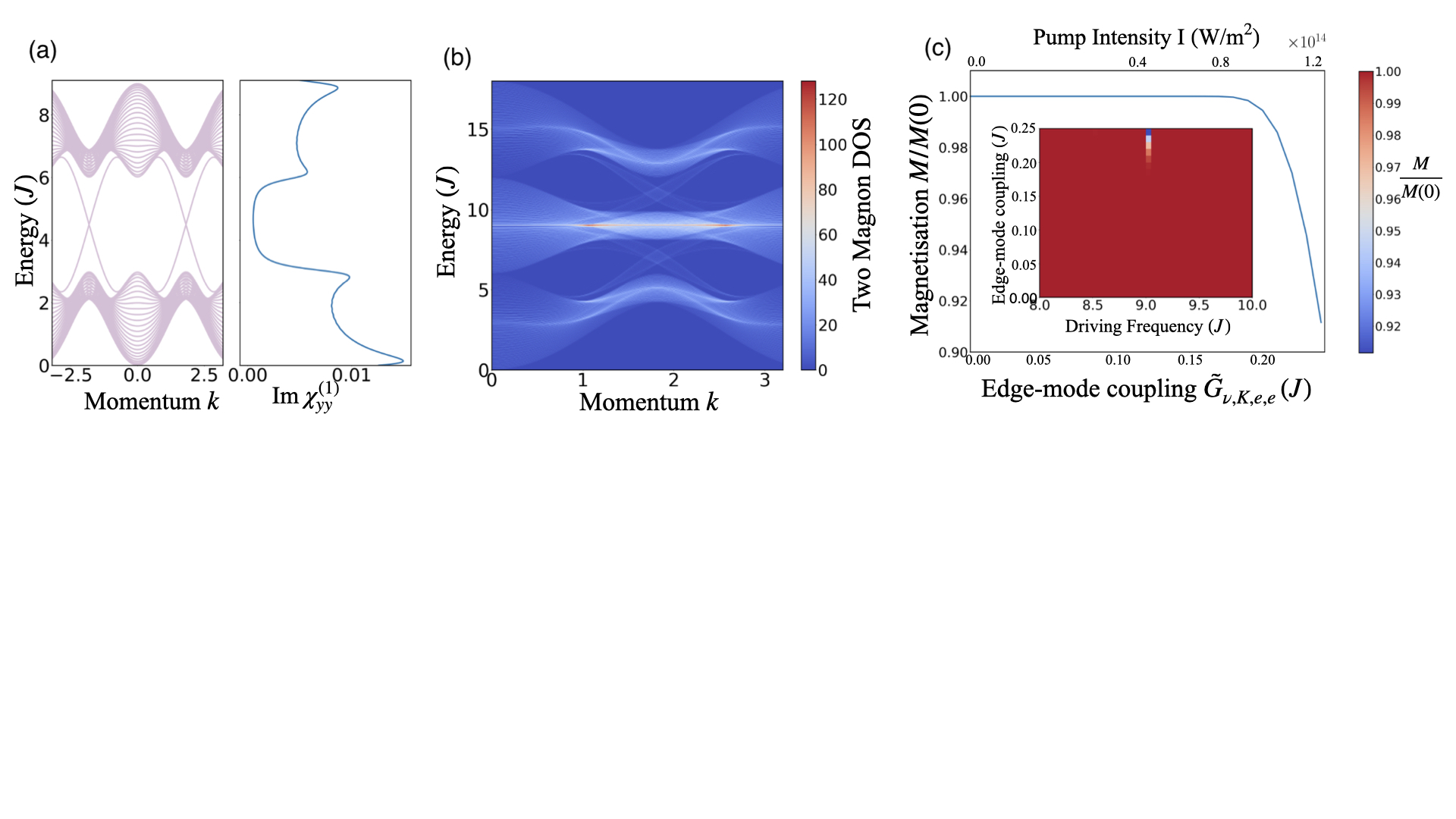}
        \captionsetup{width=0.95\linewidth}
        \captionsetup{justification=raggedright,singlelinecheck=false}
		\caption{\textbf{Demagnetization of bulk} (a) Magnon band structure and one-magnon density of states at B=0. (b) Two-magnon density of states corresponding to Eq.~\ref{eq:TDOS}. (c) Relative bulk magnetization as a function of driving intensity for $\omega_c=9J$. The inset shows the relative bulk magnetization for varying pump frequency and pump intensity, showing that the system remains magnetized up to large driving intensities. The driving intensities are shown in the top axis, corresponding to edge mode coupling strength $\tilde{G}_{\nu,K,e,e}$ varying between $0.01-0.25J$, as shown in the bottom $x$-axis.} 
        \label{fig:demag}
	\end{center}
\end{figure}

In Fig.~\ref{fig:demag}b, we depict the two-magnon density of states, defined as 
\begin{align}
    \rho^{(2)}(\omega)=\sum_{k,s,s'}\delta\left({\omega-\omega_{k,s}\pm\omega_{-k,s'}}\right)\,.
    \label{eq:TDOS}
\end{align}
A maximum density can be seen at $\omega=9J$. This frequency corresponds to magnon edge-mode pairs $2\omega_{K,e}$ as well as to several combinations of  bulk magnon modes. Therefore, one needs to be careful when driving at  $\omega_c=9J$, since two-magnon bulk processes could have dominating effect and obscure the signal from amplified edge modes.

In order to verify this, we evaluate the demagnetization of the bulk under driving at $\omega_c=9J$. The demagnetization is quantified by the reduction of the bulk magnetization arising from the population of the bulk magnon modes. From Fig.~\ref{fig:demag}c we can see that the bulk demagnetization remains small in the parameter regime we consider in this study. This behavior can be attributed to the weaker coupling of the bulk modes to the external EM field compared to the edge modes. This result indicates the bulk magnon processes contribute weakly to the response under the driving, thus ensuring that the probe response is dominated by edge magnon modes. 

\section{\label{AppendixG}Effect of boundary conditions}
\begin{figure}[hbt!]
	\begin{center}
        \includegraphics[width=0.95\textwidth, trim={0 400 50 80},clip]{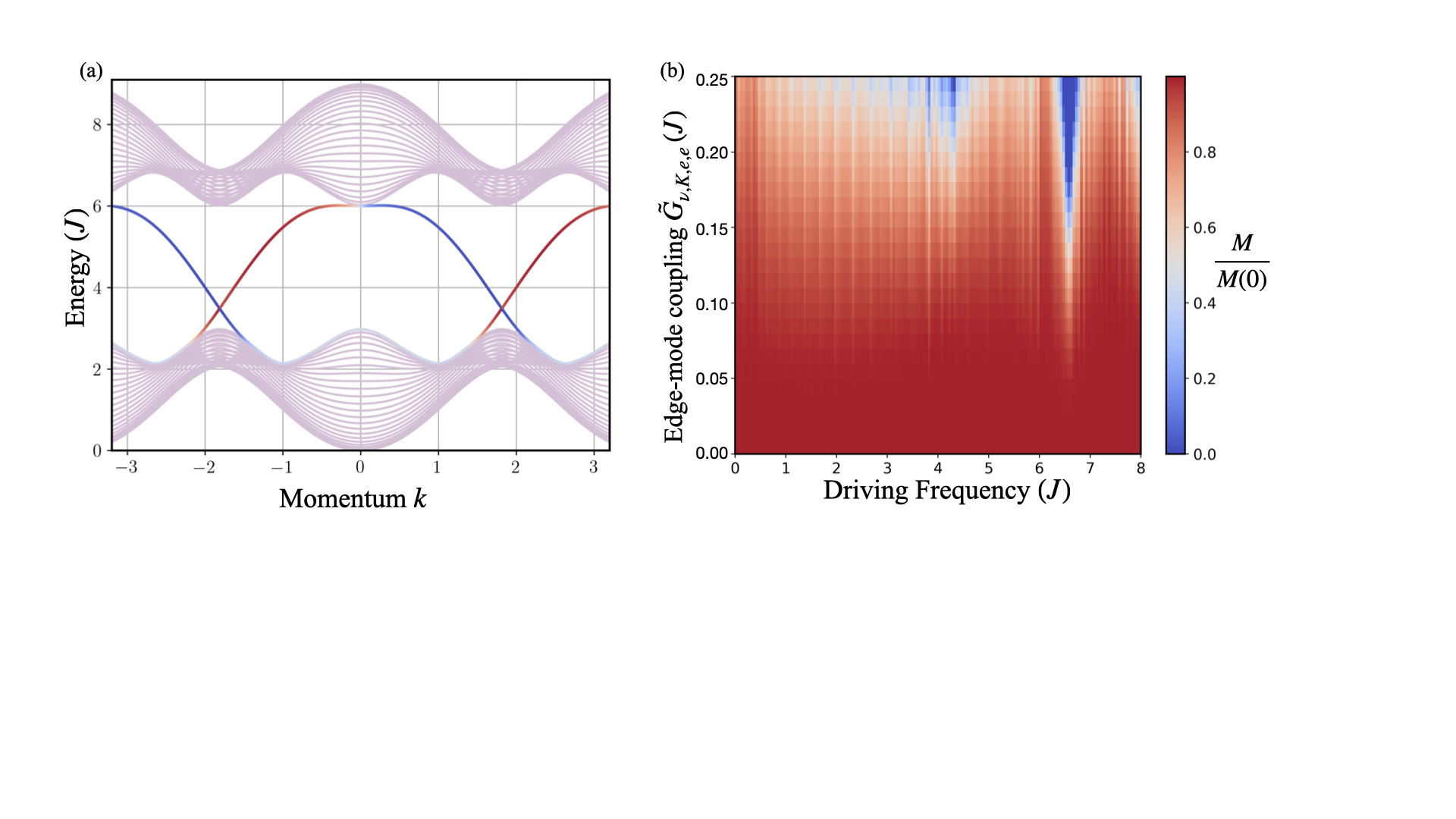}
        \captionsetup{width=0.95\linewidth}
        \captionsetup{justification=raggedright,singlelinecheck=false}
		\caption{\textbf{Zigzag Boundary conditions} We consider a zigzag boundary condition on the ribbon geometry. (a) The band structure is modified with the change of the boundary condition. The edge modes are now closer to the lower bulk band. (b) Magnetization of the bulk as a function of edge-mode coupling strength $\tilde{G}_{\nu,K,e,e}$ and frequency. The edge-mode coupling strength is directly controlled by the intensity of the drive as shown in SM~\ref{AppendixD} . A dip in the magnetization can be seen, indicating demagnetization due to increased magnon population in the bulk modes.} 
        \label{fig:Zigzag_BC}
	\end{center}
\end{figure}

In the main text we considered a bearded boundary condition on the ribbon geometry (Fig.~\ref{fig:Main}). Here, we present the result for a zigzag boundary condition. In this geometry, the edge modes lie closer to the bulk band, making it difficult to probe the edge while ensuring bulk stability. The linewidth of the pump pulse has to be smaller than 0.3J, in order to ensure excitation of only the edge mode. However, for higher incident intensity of pump laser, we have seen the edge mode splitting increases and would become comparable to the energy difference between the edge mode and the lower bulk band. Thus, the choice of a pump laser at higher intensity which resolves the bulk band and edge mode without resolving the edge mode splitting is difficult to achieve. Thus, the pump intensity is more limited as compared to the bearded geometry. From Fig.~\ref{fig:Zigzag_BC}(b), we also see that the TMI demagnetizes at higher pump intensities close to the resonant frequency $2\omega_{K,e}$. This is due to the coupling between a bulk magnon mode lying close to the edge and another bulk mode. This leads to enhanced magnon population in the bulk and a subsequent demagnetization effect. This effect occurs close to the resonant frequency, making it difficult to resolve the edge mode before destabilizing the magnetic order. This can be avoided by applying a strong magnetic field along the $z$-direction. However, the strength of the applied field required for resolving the edge modes can be expected to be larger than that required for a sample with bearded boundary condition. These constraints also affect the near-field cavity transmission measurement setup, where a higher cavity finesse and quality factor are required.
\section{\label{AppendixI}Effect of Circularly Polarized Light}
In the main text, we considered linearly polarized incident light along the finite length of the ribbon geometry. For a linearly polarized light, the polarization vector is real i.e., $\epsilon^*_{\textbf{p},\xi}=\epsilon_{\textbf{p},\xi}$. This does not hold for a circularly polarized incident light and it is convenient to use a circular basis to represent the polarization. The circular basis is related to the linear basis by a unitary transformation such that the vector fields remain invariant. For an EM wave propagating in the $z$-direction, the polarization of the field can be mapped onto the circular basis $\epsilon_{\pm}$, where $\epsilon_{\pm}=\frac{1}{\sqrt{2}}(\epsilon_{x}\pm i\epsilon_{y})$ corresponds to right and left-circular polarization and $\epsilon_{\pm}^*=\epsilon_{\mp}$. We can explicitly write the quantized EM field in the circular basis as 
\begin{align}
     \hat{\mathbf{E}}(r,t)=& \sum_{\textbf{p}}\mathcal{E}_{\textbf{p}}\Big[\hat{c}_{\textbf{p},+} (t) e^{i\textbf{p}.\textbf{r}} \epsilon_{\textbf{p},+}+\hat{c}_{\textbf{p},-} (t) e^{i\textbf{p}.\textbf{r}} \epsilon_{\textbf{p},-}+\hat{c}^{\dagger}_{\textbf{p},+} (t) e^{-i\textbf{p}.\textbf{r}} \epsilon^*_{\textbf{p},+}+\hat{c}^{\dagger}_{\textbf{p},- } (t) e^{-i\textbf{p}.\textbf{r}} \epsilon^*_{\textbf{p},-}\Big]\,,\nonumber\\
     \hat{\mathbf{B}}(r,t)=& i\sum_{\textbf{p}}\mathcal{B}_{\textbf{p}} \Big[\hat{c}_{\textbf{p},+}(t) e^{i\textbf{p}.\textbf{r}}\epsilon_{\textbf{p},+}+\hat{c}_{\textbf{p},-}(t) e^{i\textbf{p}.\textbf{r}}\epsilon_{\textbf{p},-}-\hat{c}^{\dagger}_{\textbf{p},+} (t) e^{-i\textbf{p}.\textbf{r}} \epsilon^*_{\textbf{p},+}-\hat{c}^{\dagger}_{\textbf{p},-} (t) e^{-i\textbf{p}.\textbf{r}} \epsilon^*_{\textbf{p},-}\Big]\,,
\end{align}
where $\hat{c}_{\textbf{p},\pm}=\frac{1}{\sqrt{2}}(\hat{c}_{\textbf{p},x}\mp i\hat{c}_{\textbf{p},y})$ denotes the annihilation (creation) of the left and right circularly polarized photons.

In this circular basis, the magnetic moment can be expressed as
\begin{align}
    \hat{M}^+=&\sqrt{2NS}g\mu_B\sum_{s}\pi_{0,s}\hat{\alpha}_{0,s}\,, \nonumber\\
     \hat{M}^-=&\sqrt{2NS}g\mu_B\sum_{s}\pi^{*}_{0,s}\hat{\alpha}^\dagger_{0,s}\,,
\end{align}
such that the magnetic moment in the plane perpendicular to the quantization axis can be expressed as $\hat{\textbf{M}}^{\perp}=\hat{M}^+\epsilon^++\hat{M}^-\epsilon^-$. Here, we have used $\pi_{0,s}=\pi^x_{0,s}+i\pi^y_{0,s}$. The interaction Hamiltonian in the circular basis is $\hat{H}_{int}=-\sum_{\xi,\xi '=\pm}\hat{B}^\xi\hat{M}^{\xi '}\epsilon_\xi \cdot \epsilon_{\xi '}$ or equivalently can be written as $\hat{H}_{int}=-\hat{B}^+\hat{M}^--\hat{B}^-\hat{M}^+$, where we have used the orthogonality of the unit vectors $\epsilon_\pm \cdot \epsilon_\pm=0$ and $\epsilon_+\cdot \epsilon_-=1$. For right circularly polarized incident light, the magnetic contribution to the interaction Hamiltonian is\begin{align}
    \hat{H}_{\text{int}}^{M}=-\hbar\sum_{s}\left(G^{M-}_{\pi,0,s}\hat{\alpha}_{o,s}^\dagger\hat{c}_++G_{\pi,0,s}^{M-*}\hat{\alpha}_{o,s}\hat{c}_+^\dagger\right)\,,
\end{align}
where $G_{\pi,0,s}^{M-}=i\sqrt{2NS}\sqrt{\frac{\hbar\omega_k \mu_0}{2 V}}g\mu_B\pi^*_{0,s}$ is the coupling strength per photon. Analogously, for left circularly polarized light \begin{align}
    \hat{H}_{\text{int}}^{M}=-\hbar\sum_{s}\left(G^{M+}_{\pi,0,s}\hat{\alpha}_{o,s}\hat{c}_-+G_{\pi,0,s}^{M+*}\hat{\alpha}_{o,s}^\dagger\hat{c}_-^\dagger\right)\,,
\end{align}
with $G_{\pi,0,s}^{M+}=i\sqrt{2NS}\sqrt{\frac{\hbar\omega_k \mu_0}{2 V}}g\mu_B\pi_{0,s}$. However, these processes are non-energy conserving.

For the coupling to the electric field, the leading contribution to the electric dipole moment can be written in the circular basis as 
\begin{align}
    \hat{P}_{k=0}^{(1)+}&=\sum_{s}\gamma^{+}_{0,s}\hat{\alpha}_{o,s}+\gamma^{-*}_{0,s}\hat{\alpha}^\dagger_{0,s}\,, \\ \nonumber
    \hat{P}_{k=0}^{(1)-}&=\sum_{s}\gamma^{-}_{0,s}\hat{\alpha}_{o,s}+\gamma^{+*}_{0,s}\hat{\alpha}^\dagger_{0,s}\,,
\end{align}
where the coefficients are now projected onto the circular basis $\gamma_{0,s}^{\pm}=\boldsymbol{\gamma}_{0,s}.\epsilon_{\pm}$.  One can explicitly write the terms in circular basis as 
\begin{align}
    \gamma^{+}_{0,s}= \gamma^{x}_{0,s}+i \gamma^{y}_{0,s}\,, \\ \nonumber
      \gamma^{+*}_{0,s}= \gamma^{x*}_{0,s}-i \gamma^{y*}_{0,s}\,, \\ \nonumber
        \gamma^{-}_{0,s}= \gamma^{x}_{0,s}-i \gamma^{y}_{0,s}\,, \\ \nonumber
      \gamma^{-*}_{0,s}= \gamma^{x*}_{0,s}+i \gamma^{y*}_{0,s}\,. 
\end{align}

\begin{figure}
	\begin{center}
        \includegraphics[width = 0.95 \textwidth]{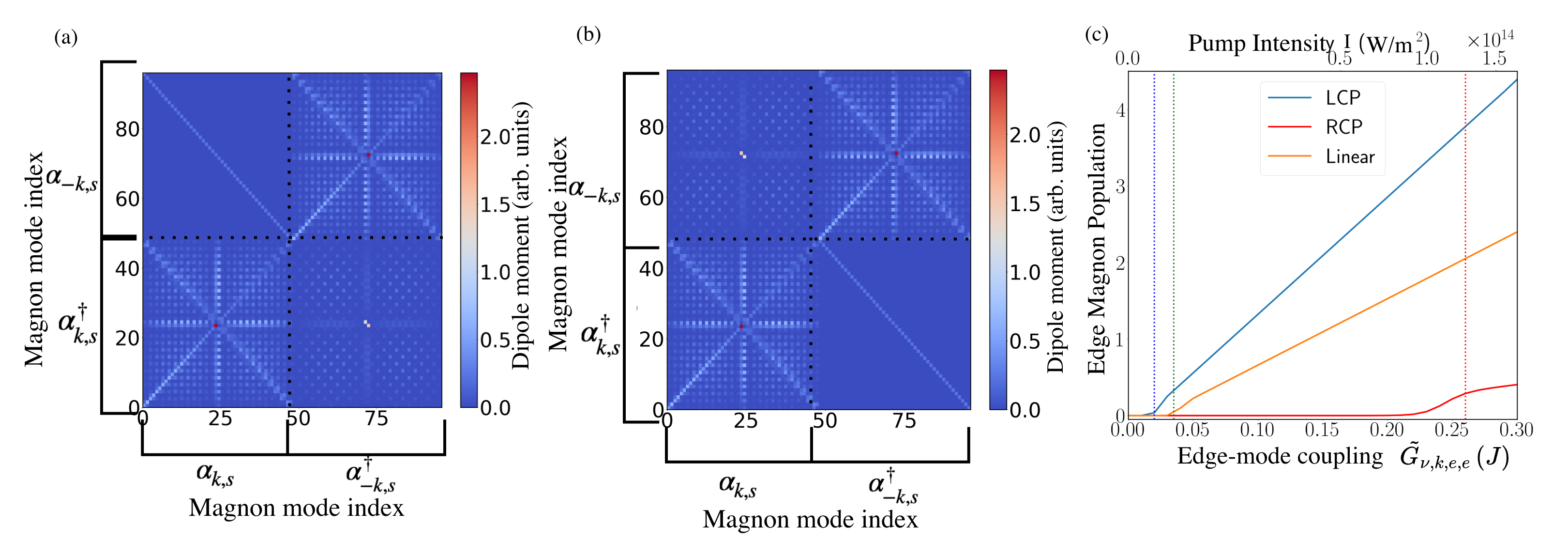}
    \captionsetup{width=0.95\linewidth}
        \captionsetup{justification=raggedright,singlelinecheck=false}
		\caption{\textbf{Electric dipole moment under circularly polarized light} The interaction between circularly polarized light and the FM TMI has different coupling strength depending on the crystal momenta $k$ and the magnon modes involved. Here we show the relative coupling strengths for magnons coupling to a circularly polarized EM field at $K$-point. Coupling strengths in matrix representation  for (a) left and (b) right circularly polarized light. In the top panel we show the magnon mode index and the corresponding frequency at the $K$-point. The index $23,24,71,72$ correspond to the edge modes. Beside the axes of the plot the basis of the matrix has been indicated. The first and fourth quadrant give coupling strengths due to scattering processes $\hat{\alpha}_{k,s}^\dagger \hat{\alpha}_{k,s'}$ and $\hat{\alpha}_{-k,s} \hat{\alpha}^\dagger_{-k,s'}$. The second and third quadrant indicate magnon pair creation $\hat{\alpha}_{k,s}^\dagger \hat{\alpha}^\dagger_{-k,s'}$ and annihilation $\hat{\alpha}_{k,s}\hat{\alpha}_{-k,s'}$ , respectively. In  (c), we show the threshold pump intensity required based on the polarization of the incident electric field. The $x$-axis shows edge mode coupling strength $\tilde{G}_{\nu,K,e,e}$ varying between $0.01-0.3J$. The corresponding physical intensity in $\text{W}/\text{m}^2$ is shown in the top axis.}
        \label{fig:Circular_polarization_2}
	\end{center}
\end{figure}

The next order terms in the magnon expansion of the electric dipole moment can be similarly projected onto the circular basis as
\begin{align}
\hat{P}_{k}^{(2)+}&=\sum_{s,s'}\mu^{+}_{k,ss'}\hat{\alpha}_{k,s}^{\dagger}\hat{\alpha}_{k,s'}+\nu^{+}_{k,ss'}\hat{\alpha}_{k,s}\hat{\alpha}_{-k,s'}+\mu^{-*}_{k,ss'}\hat{\alpha}_{k,s}\hat{\alpha}^{\dagger}_{k,s'}+\nu^{-*}_{k,ss'}\hat{\alpha}_{k,s}^\dagger\hat{\alpha}^\dagger_{-k,s'} \,, \\ \nonumber
\hat{P}_{k}^{(2)-}&=\sum_{s,s'}\mu^{-}_{k,ss'}\hat{\alpha}_{k,s}^{\dagger}\hat{\alpha}_{k,s'}+\nu^{-}_{k,ss'}\hat{\alpha}_{k,s}\hat{\alpha}_{-k,s'}+\mu^{+*}_{k,ss'}\hat{\alpha}_{k,s}\hat{\alpha}^{\dagger}_{k,s'}+\nu^{+*}_{k,ss'}\hat{\alpha}_{k,s}^\dagger\hat{\alpha}^\dagger_{-k,s'} \,,
\end{align}
where $\Lambda_{k,s,s'}^{\pm}=\boldsymbol{\Lambda}_{k,s,s'}.\epsilon_{\pm} $ for $\boldsymbol{\Lambda}=\{\boldsymbol{\mu},\boldsymbol{\nu}\}$ are the two magnon expansion coefficients in the circular basis. For a right circularly polarized incident light along $\epsilon_+$, the interaction between the electric field and magnons can be described by 
\begin{align}
\hat{H}^{E}_{\text{int}}&=-\hbar\sum_{s}\left(G_{\gamma,0,s}^{E-}\hat{\alpha}_{0,s}\hat{c}_++G_{\gamma,0,s}^{E+*}\hat{\alpha}^\dagger_{0,s}\hat{c}_++G_{\gamma,0,s}^{E+}\hat{\alpha}_{0,s}^\dagger\hat{c}_+^\dagger+G_{\gamma,0,s}^{E-*}\hat{\alpha}_{0,s}\hat{c}_+^\dagger\right) \nonumber \\
    &-\hbar\sum_{k,s,s'} \left(G_{\mu,k,s,s'}^{E-}\hat{\alpha}_{k,s}^{\dagger}\hat{\alpha}_{k,s'}+G_{\nu,k,s,s'}^{E-}\hat{\alpha}_{k,s}\hat{\alpha}_{-k,s'}+G_{\mu,k,s,s'}^{E+*}\hat{\alpha}_{k,s}\hat{\alpha}^\dagger_{k,s'}+G_{\nu,k,s,s'}^{E+*}\hat{\alpha}^\dagger_{k,s}\hat{\alpha}^\dagger_{-k,s'}\right)\hat{c}_+\\ \nonumber
    &-\hbar\sum_{k,s,s'} \left(G_{\mu,k,s,s'}^{E-*}\hat{\alpha}_{k,s}^{\dagger}\hat{\alpha}_{k,s'}+G_{\nu,k,s,s'}^{E-*}\hat{\alpha}_{k,s}\hat{\alpha}_{-k,s'}+G_{\mu,k,s,s'}^{E+}\hat{\alpha}_{k,s}\hat{\alpha}^\dagger_{k,s'}+G_{\nu,k,s,s'}^{E+}\hat{\alpha}^\dagger_{k,s}\hat{\alpha}^\dagger_{-k,s'}\right)\hat{c}^\dagger_+\,,
\end{align}
where $G^{E\pm}_{\Lambda,k,s,s'}=\frac{1}{\hbar}\sqrt{\frac{\hbar\omega_c}{2 \epsilon_0 V}}\Lambda^{\pm}_{k,s,s'}$ is the coupling strength per photon to the electric field. Similarly, the interaction Hamiltonian for an incident left circularly polarized light can be obtained by exchanging the indices $+ \leftrightarrow-$. The mode dependence of the couplings in circular $\Lambda^{\pm}_{k,s,s'}$ is shown in  Fig.~\ref{fig:Circular_polarization_2}. The asymmetry of the electric dipole moment contribution in the left and right circular components shows that the electric dipole moment is elliptically polarized. 
 
 Now we turn to analyze the angular momentum (AM) conservation in the individual processes. The magnons of the FM TMI are RCP and carry angular momentum $-\hbar$. The photon fields have spin angular momentum which can be obtained from the EM fields as~\cite{Circular_qunatization}
\begin{align}
    \hat{S}=\epsilon_0\int d^3r \Big(\hat{\textbf{E}}(r)\times \hat{\textbf{A}}(r)\Big).
\end{align}
For circularly polarized light, we explicitly calculate the spin-angular momentum of the photon field using the quantized form of the vector field $\hat{\textbf{A}}(r)$ and $\hat{\textbf{E}}(r)$ from above and obtain
\begin{align}
    \hat{S}=\hbar\sum_\textbf{p} \Big[ \hat{c}_{\textbf{p},-}^\dagger\hat{c}_{\textbf{p},-}-\hat{c}_{\textbf{p},+}^\dagger\hat{c}_{\textbf{p},+} \Big].
\end{align}
This shows that the photons with polarization along $\epsilon_-$ have angular momentum $\hbar$, correspond to left circular polarization (LCP) and those with polarization along $\epsilon_+$ have angular momentum $-\hbar$, corresponding to a right circularly polarization. For linearly polarized light, the photon fields are not eigenvectors of the angular momentum operator.

The coupling between circularly polarized light and magnons can lead to processes which can be divided into three categories: a) one magnon coupled to one photon, b) magnon-pair creation/ annihilation mediated by photon absorption/creation and c) scattering between magnons due to a circularly polarized photon. Suppose the change in AM due to magnons and photons be $\Delta S_m$ and $\Delta S_p$, respectively. For a system with continuous rotational symmetry, AM conservation requires $\Delta S=\Delta S_m+\Delta S_p=0$. However, we are interested in a honeycomb lattice with $C_3$ symmetry, which is imposes conservation of AM modulo $3\hbar$~\cite{CrI3_observation}. 

For the first category, the allowed processes include creation of a magnon by absorption of an RCP photon or vice-versa, where a photon with AM $-\hbar$ is annihilated creating a magnon with $-\hbar$, thereby leading to zero net angular momentum.  For magnon-pair creation(annihilation) processes, the change in AM due to pair creation(annihilation) is $\Delta S_m=-2\hbar(+2\hbar)$.  For a system with continuous rotational symmetry, the change in AM required due to the photon field to ensure conservation of AM is $\Delta S_p= \pm2\hbar$, which can be achieved either by using structured light or through scattering between circularly polarized light. But due to the honeycomb lattice symmetry, the selection rules allow creation of magnon pairs by absorption of an LCP photon or via emission of RCP photon, which lead to conservation of AM modulo $3\hbar$. The latter process is a non-energy conserving process and can be neglected under RWA. Similarly, magnon pairs can be absorbed to create LCP photons. The third process involves scattering between magnons due to a circularly polarized photon, in which due to the identical handedness of the magnons, $\Delta S_m=0$. In this case, $\Delta S_p=\pm\hbar$, which leads to $\Delta S=\Delta S_m+\Delta S_p\neq0(mod\ 3\hbar)$, violating the conservation of AM. However, the magnon scattering processes originate from a chiral term in the electric dipole moment expansion i.e., $\hat{\textbf{S}}_i\times\hat{\textbf{S}}_j$. This indicates the possibility of a non-zero orbital angular moment(OAM) for the magnons~\cite{OAM_magnon}. Thus the AM conservation should be checked including the contribution of OAM. Note that, moreover, the previous arguments hold for the bulk. Since the ribbon geometry breaks rotational symmetry, angular momentum is not necessarily conserved when explicitly considering the finite geometry.

We study now the dynamics of the magnons under a classical circularly polarized incident light. As discussed in Sec.~\ref{sec:Dynamics}, the effective coupling to the drive is controlled by the intensity of the incident field. Upon simulating the dynamics, we observe that for LCP light, there is amplification of magnon edge modes and the threshold pump intensity required for this amplification is lower than that required for linearly polarized light. However, for large pump intensities, off-resonant processes also start contributing and in this regime, amplification can be obtained with RCP light as well.

\end{document}